\documentclass[10pt,compsocconf,conference]{IEEEtran}
\IEEEoverridecommandlockouts

\usepackage{cite}
\usepackage{lipsum}
\usepackage{multirow}
\usepackage{amsmath}
\usepackage{amssymb}
\usepackage{amsfonts}
\usepackage{algorithmic}
\usepackage{graphicx}
\usepackage{textcomp}
\usepackage{float}
\usepackage{blindtext}
\usepackage{xcolor}
\usepackage{caption}
\usepackage{array}
\usepackage{booktabs}
\newcolumntype{C}[1]{>{\centering\let\newline\\\arraybackslash\hspace{0pt}}m{#1}}
\newcommand{\ApproxSign}{\raise.17ex\hbox{$\scriptstyle\sim$}}

\def\BibTeX{{\rm B\kern-.05em{\sc i\kern-.025em b}\kern-.08em
    T\kern-.1667em\lower.7ex\hbox{E}\kern-.125emX}}

\newcommand\blfootnote[1]{%
  \begingroup
  \renewcommand\thefootnote{}\footnote{#1}%
  \addtocounter{footnote}{-1}%
  \endgroup
}
\usepackage{enumitem}
\setlist[enumerate]{leftmargin=*}
\begin{document}

\title{DRACO: Co-Optimizing Hardware Utilization, and Performance of DNNs on Systolic Accelerator\\
{}
}

\author{\IEEEauthorblockN{Nandan~Kumar~Jha$^\star$, Shreyas~Ravishankar$^\dagger$, Sparsh~Mittal$^\ddagger$, Arvind~Kaushik$^\S$, Dipan~Mandal$^\P$, Mahesh~Chandra$^\S$}
\IEEEauthorblockA{
$^\star$IIT Hyderabad, $^\dagger$BITS Pilani Hyderabad, $^\ddagger$IIT Roorkee, $^\S$NXP Semiconductors, $^\P$Intel Labs, India. \\
Email: cs17mtech11010@iith.ac.in, f20160180@hyderabad.bits-pilani.ac.in, sparshfec@iitr.ac.in,\\arvind.kaushik@nxp.com, dipan.mandal@intel.com, mahesh.chandra\_1@nxp.com
}}

\maketitle

\begin{abstract}
The number of processing elements (PEs) in a fixed-sized systolic accelerator is well matched for large and compute-bound DNNs; whereas, memory-bound DNNs suffer from PE underutilization and fail to achieve peak performance and energy efficiency.  To mitigate this, specialized dataflow and/or micro-architectural techniques have been proposed. However, due to the longer development cycle and the rapid pace of evolution in the deep learning fields,  these hardware-based solutions can be obsolete and ineffective in dealing with PE underutilization for state-of-the-art DNNs.  In this work, we address the challenge of PE underutilization at the algorithm front and propose data reuse aware co-optimization (DRACO). This improves the PE utilization of memory-bound DNNs without any additional need for dataflow/micro-architecture modifications. Furthermore, unlike the previous co-optimization methods, DRACO not only maximizes performance and energy efficiency but also improves the predictive performance of DNNs. To the best of our knowledge, DRACO is the first work that resolves the resource underutilization challenge at the algorithm level and demonstrates a trade-off between computational efficiency, PE utilization, and predictive performance of DNN. Compared to the state-of-the-art row stationary dataflow, DRACO achieves 41.8\% and 42.6\% improvement in average PE utilization and inference latency (respectively) with negligible loss in predictive performance in MobileNetV1 on a $64\times64$ systolic array. DRACO provides seminal insights for utilization-aware DNN design methodologies that can fully leverage the computation power of systolic array-based hardware accelerators.
\end{abstract}

\begin{IEEEkeywords}
Deep neural networks (DNNs), systolic array, PE utilization, latency, energy-efficiency. 
\end{IEEEkeywords}

\section{Introduction}
Systolic array has become mainstay in hardware-based Deep Neural Network (DNN) accelerators deployed in both edge devices \cite{2017_Chen_JSSC} and servers used in data-centers \cite{shao2019simba,jouppi2017datacenter}. The systolic array-based architectures enable local communication between neighboring processing elements (PEs) which exploits the inherent parallelism in DNN and adds one extra level in the memory hierarchy of accelerators.  This, together with the distributed control logic in systolic architecture, offers an efficient solution for DNN's processing\blfootnote{Support for this work was provided by Semiconductor Research Corporation.} \cite{chen2016eyeriss}.

Concurrently, on the DNN algorithm side, depthwise convolution (DWConv) \cite{Howard2017MobileNetsEC,2018_CVPR_Sandler,Zhang_2018_CVPR} has been devised to reduce the number of computations (measured in terms of multiply-accumulation or MAC operations) and parameters. Unlike standard convolution (SConv), where entire channels in a filter are convolved with all input feature maps (ifmaps) and produce one output feature map (ofmap), DWConv allows one filter-channel to convolve with only one ifmap. Consequently, DWConv exhibits low data reuse and parallelism, which makes DNNs with DWConv as memory-bandwidth bound  \cite{park2018deep}. The performance and energy efficiency of systolic array-based DNN accelerators rely heavily on certain pre-determined dimensions of convolution (conv) layers and/or feature maps, which helps exploiting high parallelism in large and compute-bound DNNs \cite{chen2019eyeriss}. However, the reduced shape and size of conv layers, including DWConv, in compact DNNs render most of the PEs underutilized in a fixed-sized systolic accelerator that are designed for large and compute-bound DNNs.

\begin{figure}[htbp]
\centering
\includegraphics[scale=0.2]{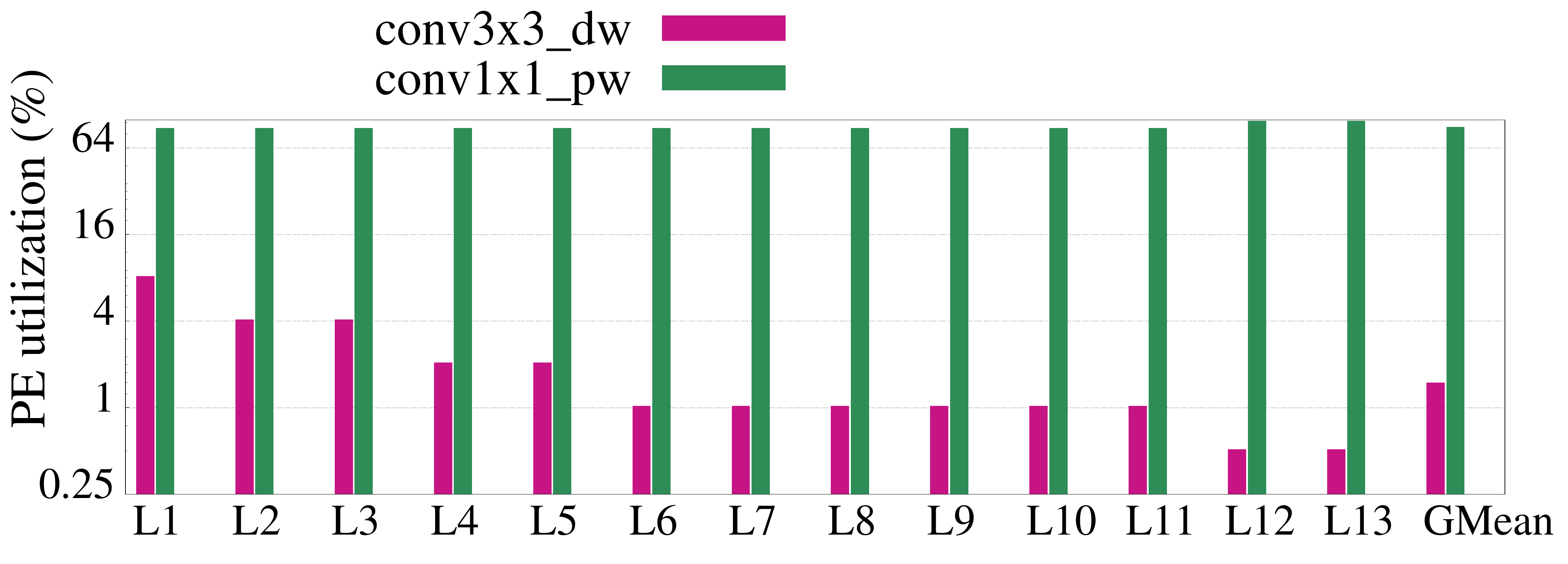}  
\caption{Layer-wise PE utilization in MobileNetV1}
\label{fig:LayerWisePEutil}
\end{figure}

The low PE utilization in systolic array poses three-fold challenges. First, it increases the stall cycles and hence prediction latency, which prohibits inference in real-time. Second, the accelerator does not attain its peak performance and energy-efficiency. Third, the PE underutilization becomes more significant in the larger array sizes and exacerbates the scalability challenge.  Fig. \ref{fig:LayerWisePEutil} shows the experimental results (refer Section \ref{sec:ExperimentalResults} for details on  experimental setup) for layer-wise PE utilization  on a 64$\times$64 systolic array for MobileNetV1 (MV1). Evidently, the utilization is very high for $1\times1$ pointwise conv.  However, utilization of the $3\times3$ DWConv  is low in initial layers (only {\bf $\approx$4\%}) and decreases further in deeper layers due to the reduced size of fmaps. The {\em key observation} here is, PE utilization depends on both the number of channels in groups and the spatial size of fmaps.

To deal with the aforementioned challenges  in DWConv, specialized micro-architecture \cite{shao2019simba,chen2019eyeriss} and dataflow techniques \cite{2019_DATE_Wu,2019_ASPDAC_Liu} have been proposed. From the general perception in the realm of co-design, a hardware-based solution yields lower return  on investment due to prolonged development cycle and lack of flexibility, hence it is used as a last resort. Therefore,  we  resolve the issue of PE underutilization at the algorithm level and provide  generic guidelines for utilization-aware DNN design. We seek to answer  {\em whether we can change the architecture of DNN itself such that memory-bound DNNs can efficiently run on the accelerators conventionally  designed for large and compute-bound DNNs without any micro-architecture/dataflow modifications?}

In this paper, we propose data reuse aware co-optimization (DRACO) where the number of channels in the groups of fmaps are set to optimize the PE utilization on systolic-array based DNN accelerators. Moreover, by merely fine-tuning the number of channels in the groups of fmaps, DRACO can easily adapt to various sizes of systolic array. Through our extensive experiments, we demonstrate that DRACO {\em  enables substantial improvement in PE utilization on a wide range of systolic array size with a minimal increase in the number of MAC operations and parameters}. We also examine the energy-efficiency implication of DRACO. We find that due to better data reuse in DRACO, the increase in computation increases the access to only local memory (register file or RF) rather than off-chip DRAM. Thus, the increase in total energy consumption is meager. {\em Interestingly},  we observe that DRACO not only preserves the accuracy of the baseline model, but it can also boost the predictive performance with an appropriate number of channels in the groups of fmaps.

DRACO is applicable to a wide range of DNNs which employ either DWConv or group convolution with a lesser number of channels per group of fmaps. Moreover, it can also be used in the lighter (mobile) version of large DNNs. For instance, in the bottleneck modules of ResNet, $3\times3$ SConv can be  replaced with $3\times3$ group convolution with an optimal number of channels per group and it can be used as a mobile version of ResNet with high PE utilization on different scales of PEs.
Our {\bf contributions} are as follows.
\begin{itemize}
\item We implement the DRACO technique to mitigate the low data reuse and PE underutilization in DWConv of memory-bound DNNs.  
\item We perform an extensive experimentation to understand the interplay of computational complexity and PE utilization with  (inference) latency. We reveal that increasing PE utilization at the expense of high computational cost negates the effect of improved utilization and results in suboptimal inference latency. 
\item We demonstrate the efficacy of DRACO by performing experiments with four different array sizes.
\item We perform accuracy experiments on imagenette \cite{Fastai_Imagenette} to show the predictive performance implication of DRACO. 
\end{itemize}
 
\section{Related Work} \label{sec:Background}

In Eyeriss v2 \cite{chen2019eyeriss}, hierarchical mesh NoC has been proposed to cater to the high bandwidth requirement for DNNs with low data reuse. Simba \cite{shao2019simba} deployed Global PE, which performs the near-memory computation for the DNN's layer with low data reuse. The authors in \cite{2017_DAC_Wei} developed an analytical model to efficiently explore the design space and maximize the resource utilization in a systolic array. Liu et al. \cite{2019_ASPDAC_Liu} propose flexible mappings for computation in conv layers to attain high PE utilization.  Similarly, Wu et al. \cite{2019_DATE_Wu} implement an optimized execution order for tiled matrix multiplication to maximize the data reuse in depthwise separable convolution. Kung et al.  \cite{2018_ICPR_Kung} employ adaptive tiling to reduce the number of tiles required for inference in sparse (unstructured) DNNs. Unlike the aforementioned ad hoc solutions, we change the architecture of DNN itself to achieve a significant gain in PE utilization with a minimum increase in computations and energy consumption. We also study the effect of the proposed algorithmic change on the predictive performance of DNNs.

\section{Proposed Approach} \label{sec:Methodology}

{\bf Notations} The spatial size of filter (kernel) and fmap are denoted as $d_k\times d_k$ and $d_f\times d_f$ respectively (Table \ref{tab:DataReuse}). For simplicity, we assume the spatial size of ifmap and ofmap are equal. $n$ and $m$ are the number of ifmaps and ofmaps, respectively. The number of ifmaps, and channels in a filter; and the number of filters, and ofmaps are equal. The number of channels (ifmaps) in a group of filters (fmaps) is denoted as $G$, and the total number of such groups in a conv layer is $\frac{m}{G}$. The number of parameters and activations (input and output together) are denoted as $\# Param$ and $\# Act$, respectively.  In Table \ref{tab:DataReuse}, data reuse  is estimated as arithmetic intensity of the layer ($DataReuse$=$\frac{\#MACs}{\#Param \; + \;\#Act}$). 
We separately calculate the data reuse for filter-weights as  $W_{reu} = \frac{\# MACs}{\# Param}$. Also, data reuse for input/output activations is   $A_{reu} = \frac{\# MACs}{\# Act}$.

\subsection{Data Reuse Bottleneck in Depthwise Convolution}

\begin{figure}[htbp]
\centering
\includegraphics[scale=0.23]{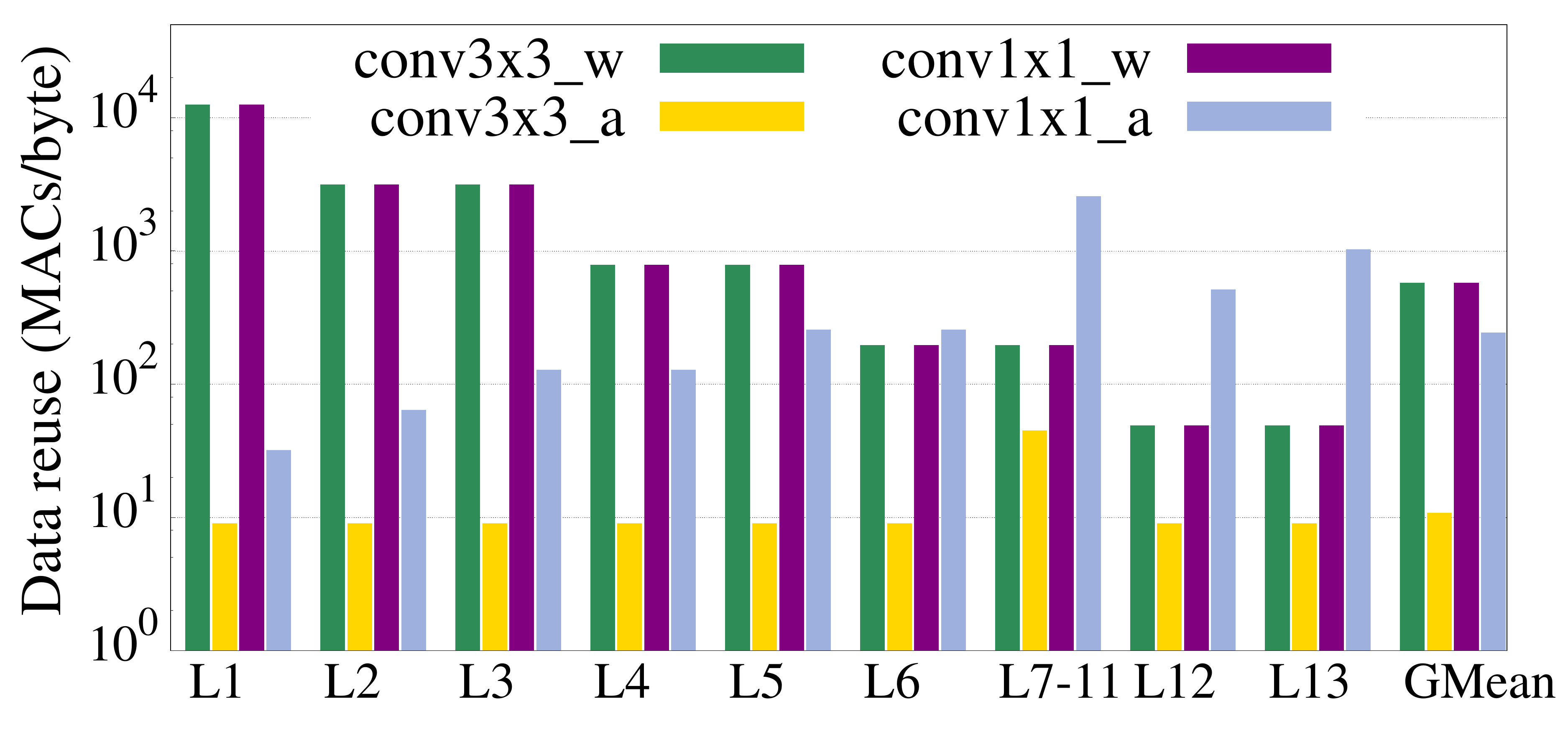}  
\caption{Weight reuse in 3$\times$3 DWConv (conv3x3\_w), and 1$\times$1 convolution (conv1x1\_w); activation reuse in 3$\times$3 DWConv (conv3x3\_a), and 1$\times$1 convolution (conv1x1\_a)}
\label{fig:DataReuseMV1}
\end{figure}

In comparison with standard convolution (SConv), DWConv ($G$=1) reduces both the number of MAC operations and parameters by a factor of $n$ (Table \ref{tab:DataReuse}) which is significant especially in deeper layers where $n$ is very high. However, the data reuse of DWConv is very low compared to SConv. As shown in Table \ref{tab:DataReuse}, the $W_{reu}$ is the same in DWConv and SConv; whereas, $A_{reu}$ of the former is reduced by a factor of $n$. For example, as illustrated in Fig. \ref{fig:DataReuseMV1}, $W_{reu}$ of $3\times3$ DWConv and $1\times1$ SConv in  MobileNetV1 are same; however, $A_{reu}$ in the former is significantly lower than the latter. Also, this disparity in $A_{reu}$ grows in deeper layers. Note that {\em $A_{reu}$ is more critical} than $W_{reu}$ because $W_{reu}$ increases with increasing input batch size, while $A_{reu}$ does not \cite{park2018deep}.

\begin{table} [htbp]
\caption{Data reuse comparison }
\label{tab:DataReuse}
\centering 
\resizebox{0.49\textwidth}{!}{
\begin{tabular}{ |c| c| c| c| } 
 \hline
 {\bf Metric} &  {\bf SConv} & {\bf DWConv} & {\bf DRACO}   \\
 \hline
 \#MACs &  $m\times n\times d_k^2\times d_f^2$ & $m\times d_k^2\times d_f^2$ & $G\times(n\times d_k^2\times d_f^2)$\\ \hline
 $\#Param$ & $m\times n\times d_k^2$  &  $m\times d_k^2$ & $G\times(n\times d_k^2)$\\ \hline
 $\#Act$ &  $(m + n)\times d_f^2$ & $(m + n)\times d_f^2$ & $(m + n)\times d_f^2$\\ \hline
  $DataReuse$  &  $\frac{m\times n\times d_k^2\times d_f^2}{m\times n\times d_k^2 + (m + n)\times d_f^2}$ &$\frac{m\times d_k^2\times d_f^2}{m\times d_k^2 + (m + n)\times d_f^2}$ & $\frac{m\times d_k^2\times d_f^2}{m\times d_k^2 + \frac{(m + n)}{G}\times d_f^2}$ \\ \hline
  $W_{reu}$ & $d_f^2$ & $d_f^2$ & $d_f^2$ \\ \hline
 $A_{reu}$ & $n\times\big(\frac{m}{m+n}\big)d_k^2$ &$\big(\frac{m}{m+n}\big)d_k^2$ & $G\times\big(\frac{n}{m+n}\big)d_k^2$ \\
\hline
\end{tabular} }

\end{table}

Dataflow techniques are selected to maximize the data reuse of a particular data type in DNN. For example,  weights and partial sum reuse are maximized in weight stationary and output stationary dataflow \cite{chen2016eyeriss}. In initial layers, $W_{reu}$ is higher and decreases in the deeper layer, whereas $A_{reu}$ is higher in deeper layers and decreases in layers towards input. Consequently, none of the dataflows can be optimal for all the layers in a DNN. In other words, due to the lower number of filter-channels (or ifmaps) in the initial layers and smaller fmap size in deeper layers, 
PE underutilization happens in the respective layers. This underutilization is further aggravated in DWConv due to the very low $A_{reu}$.

\subsection{Data Reuse Aware Co-optimization (DRACO)}

First, we employed state-of-the-art dataflow, {\em row stationary} \cite{chen2016eyeriss}, which is adaptable to different shapes and sizes of filters/ifmaps, thus, enables optimal energy-efficiency and throughput for a DNN. As illustrated in Fig. \ref{fig:LayerWisePEutil}, the PE utilization of $1\times1$ conv with row stationary dataflow is well-above of $\approx$80\% and independent of the layer's position in DNN. However, in the case of DWConv, PE utilization is substantially lower ($\leq$ 4\%) on large array size ($64\times64$) and {\em more importantly, it reduces in deeper layers}. That is, even with row stationary dataflow, when the number of channels in a group is deficient, then PE utilization depends on the spatial size of fmaps, and hence it decreases in deeper layers.

To address the shortcomings of row stationary dataflow and to enable high PE utilization in DWConv, we propose data reuse aware co-optimization (DRACO) where we increase the $G$, which in turn increases $A_{reu}$ (Table \ref{tab:DataReuse}). Since PE underutilization becomes more significant with larger array size, by fine-tuning $G$, DRACO can adapt to larger array size. Increasing $G$ also increases the number of MAC operations, but, increases the data reuse and decreases the bandwidth pressure. In other words, DRACO adds more \#MACs to a bandwidth bound DNN without increasing the bandwidth demand and substantially improves the PE utilization.

\section{Experimental Evaluation} \label{sec:ExperimentalResults}
In this section, we detail the experimental setup and elaborate on our findings. We present the insights as key takeaway messages, which can enable a better methodology for latency optimization in DNNs. 

\subsection{Experimental Setup}

{\bf Eyeriss simulator} We take Eyeriss, a systolic array-based inference accelerator for DNNs \cite{2017_Chen_JSSC}, as a baseline architecture and modify the Eyeriss simulator \cite{gao2019tangram,gao2017tetris} to perform our experiments. Table \ref{tab:ExperimentalSetup} shows the on-chip memory (global buffer and register file) sizes for different PE array sizes. 
Consistent with Chen et al. \cite{chen2016eyeriss}, the memory cost for accessing the DRAM (off-chip), global buffer (on-chip SRAM), array (inter-PE), and register file (local scratch-pad) are modeled as \{200, 6, 2, 1\}$\times10^{-12}$ joules, respectively. Since inference is latency-critical, we take batch size as one for all the experiments.

\begin{table} [htbp] 
\caption{Configuration of on-chip memories in Eyeriss simulator for different array sizes} 
\label{tab:ExperimentalSetup}\centering 
\begin{tabular}{ |c|c|c|c|c| } \hline
 {\bf Memory size} & \textbf{16x16} & \textbf{32x32}  & \textbf{64x64} & \textbf{128x128} \\ \hline 
 GBuf (KiB) &  128 & 256 & 512 & 1024  \\
 RF per PE (KiB) & 0.5 & 0.5 & 0.5 & 0.5 \\ 
 Total RF size (KiB) & 128 & 512 & 2048 & 8192 \\ \hline
 \end{tabular}  
\end{table}

{\bf Baseline model and dataset} DWConv ($G$=1) is a popular technique to reduce the computation in both manual design and automated design of DNNs \cite{tan2019efficientnet}. We use MobileNetV1 \cite{Howard2017MobileNetsEC}, widely deployed on mobile platforms, as a baseline model because all of its layers with $3\times3$ conv (except very first layer) employ DWConv. In MobileNetV1, all  $3\times3$ conv layers are followed by a $1\times1$ conv layer, and it bears a resemblance to a widely deployed building block:  bottleneck module \cite{He_2016_CVPR}. Therefore, the findings obtained for MobileNetV1 also apply to an extensive range of DNNs.

\begin{figure}[htbp]
\centering
\includegraphics[scale=0.54]{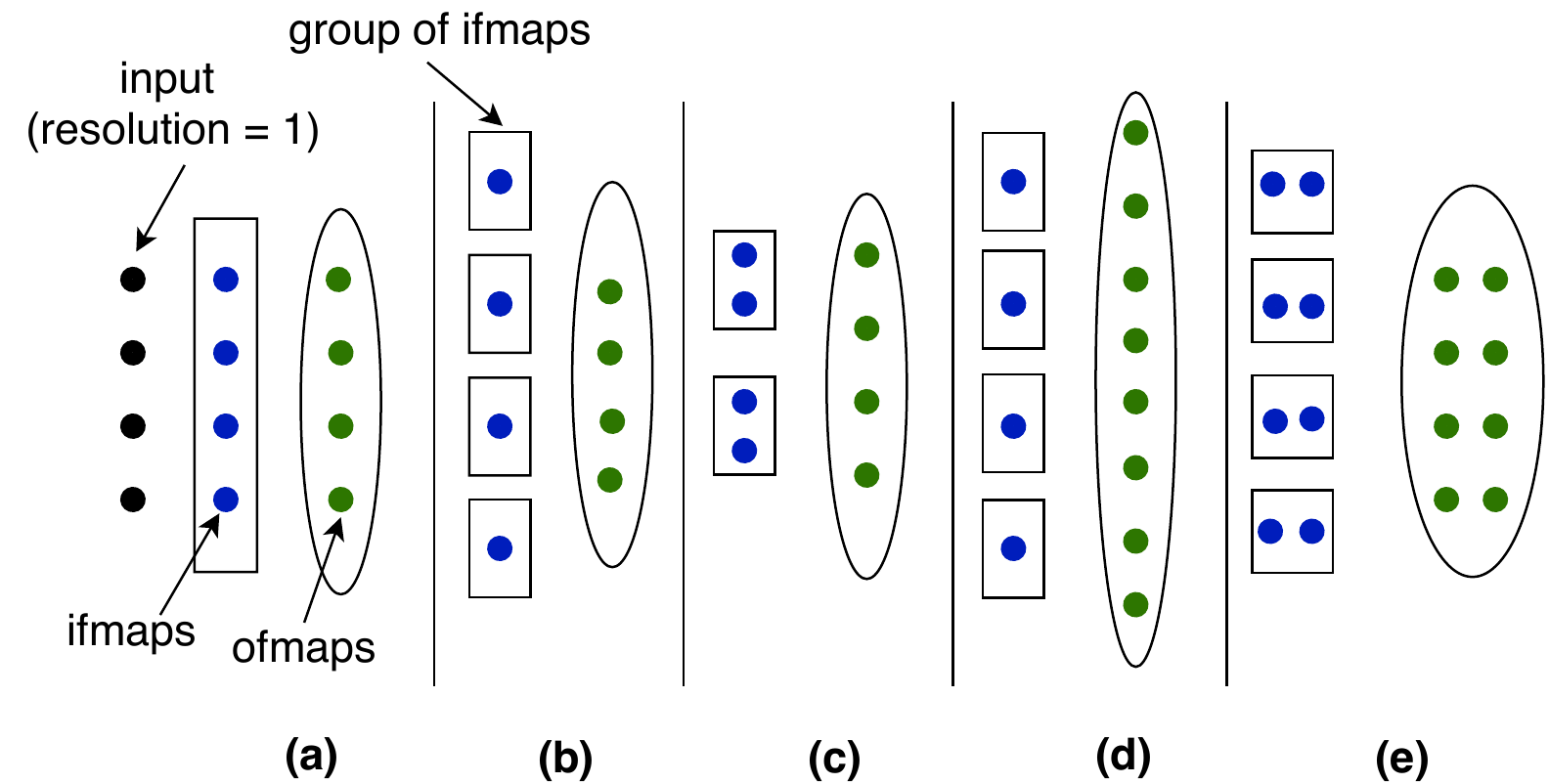}  
\caption{Comparison of (a) SConv with (b) MobileNetV1 ($\alpha$=1, $\rho$=1, $G$=1); and other MobileNetV1 variants with (c) \{$\alpha$=1, $\rho$=1, $G$=2\}, (d) \{$\alpha$=2, $\rho$=1, $G$=1\}, and (e) \{$\alpha$=1, $\rho$=2, $G$=1\}}
\label{fig:VersionsOfMV1}
\end{figure}

To  understand the interplay of compute efficiency and PE utilization with inference latency, we perform experiments on MobileNetV1 with different values of width multiplier ($\alpha$) and input resolution multiplier ($\rho$) \cite{Howard2017MobileNetsEC}. The physical significance of $\alpha$, $\rho$, and $G$ are illustrated in Fig. \ref{fig:VersionsOfMV1}. Changing $\alpha$ changes both the number of parameters and MACs whereas $\rho$ changes only the number of MACs.  As the number of channels in first $3\times3$ DWConv layer of MobileNetV1 is 32, we perform our experiments for $G$=1 to $G$=16 for $\alpha$=0.5, $G$=1 to $G$=32 for $\alpha$=1 and $G$=1 to $G$=64 for $\alpha$=2.  Table \ref{tab:MACsParamsMV1All} shows the number of MACs and parameters with different values of  $\alpha$ and $\rho$. Note that to accommodate larger fmaps in the case of MobileNetV1 with $\rho$=2, we double the on-chip memory (GBuf and RF per PE) capacity shown in Table \ref{tab:ExperimentalSetup}.

\begin{figure*}[htbp]
\centering
\includegraphics[scale=0.116]{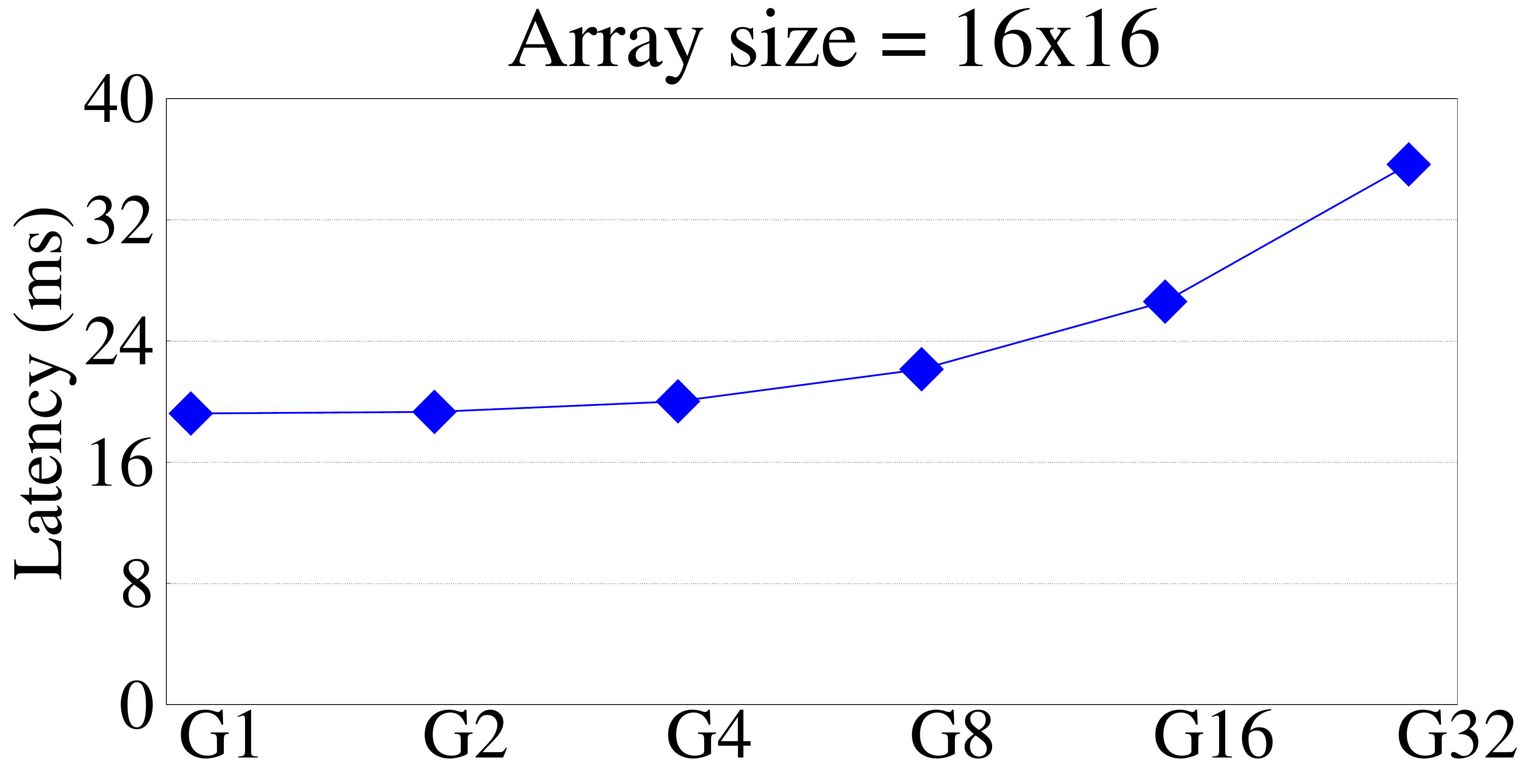}  
\includegraphics[scale=0.116]{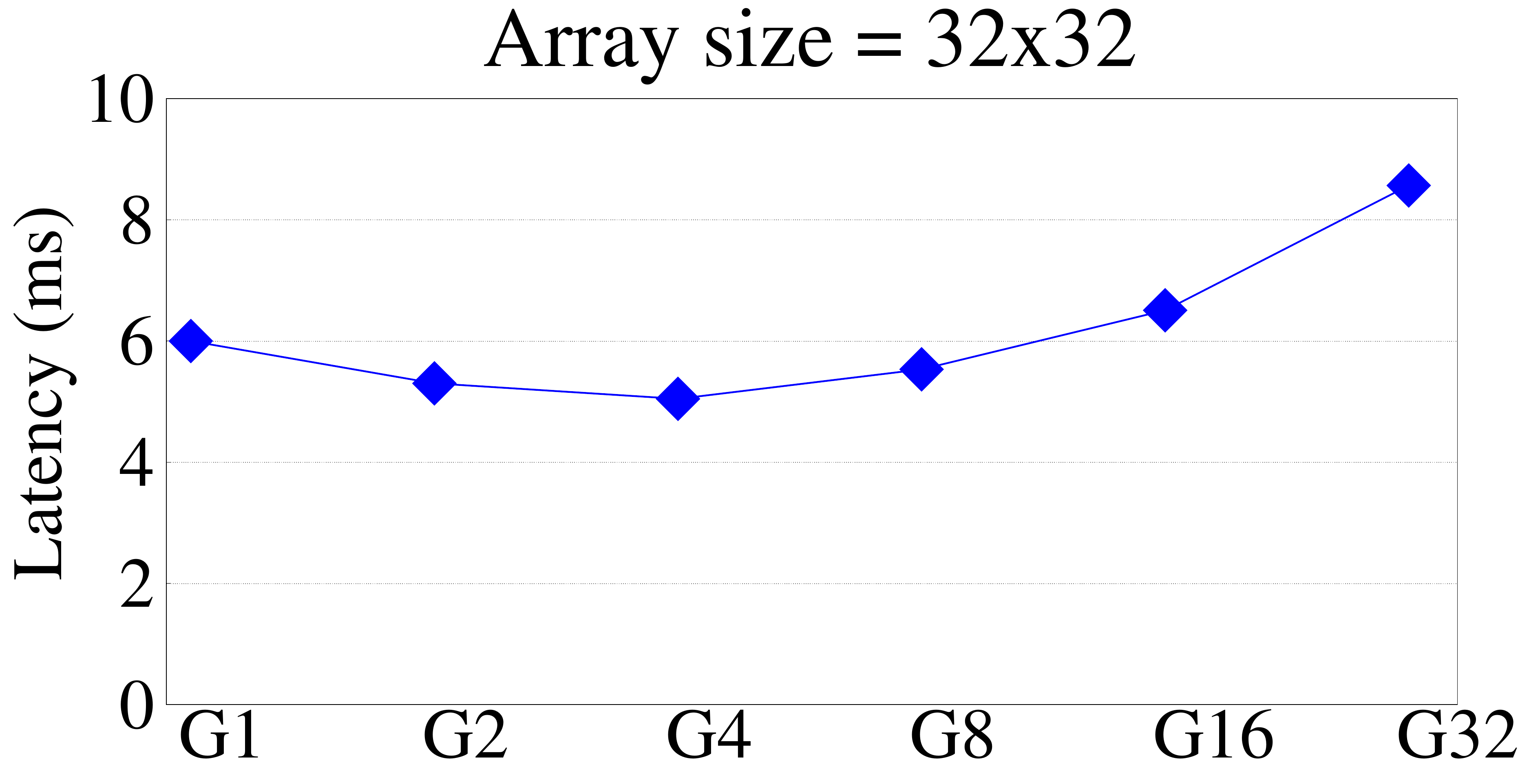} 
\includegraphics[scale=0.116]{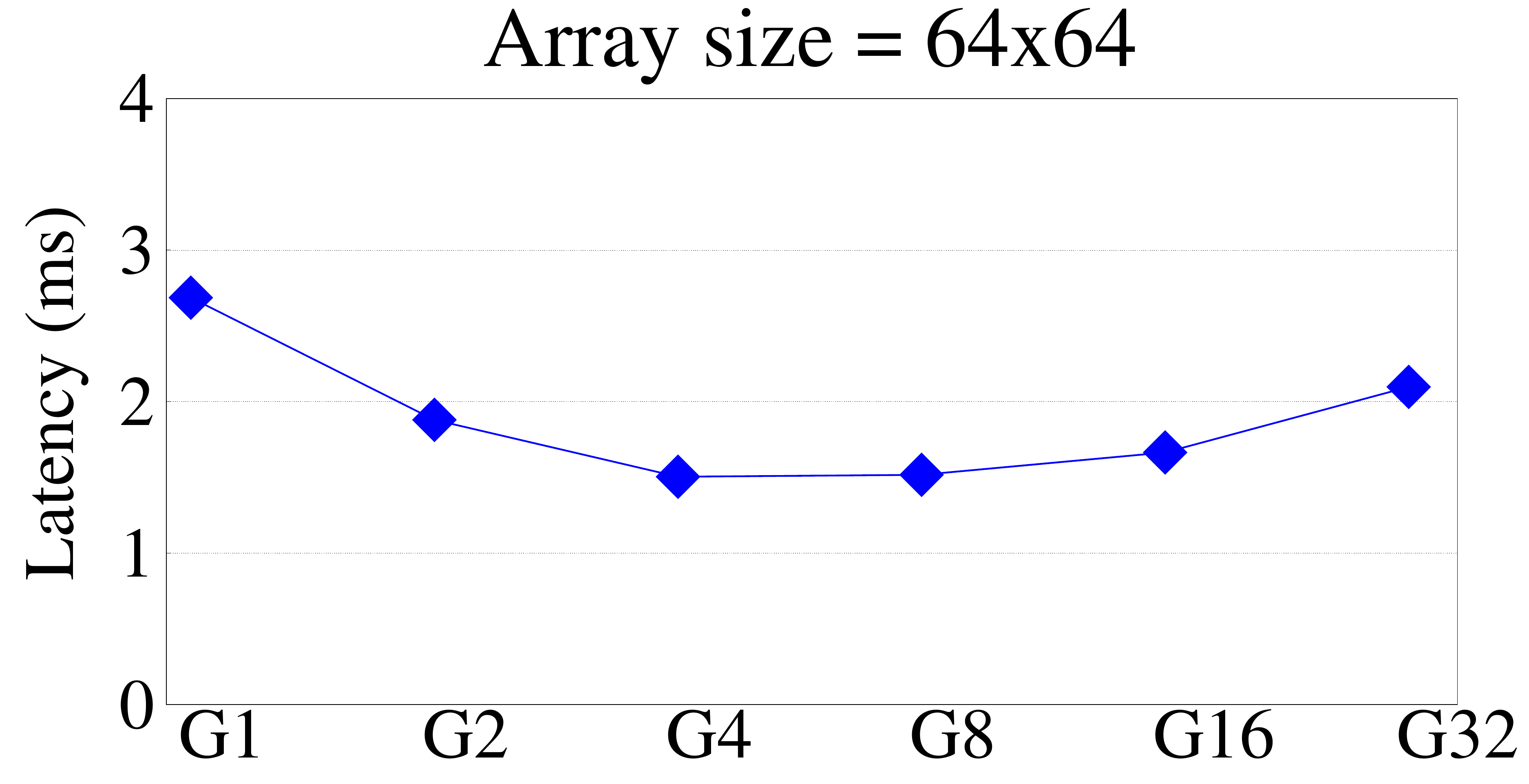}  
\includegraphics[scale=0.116]{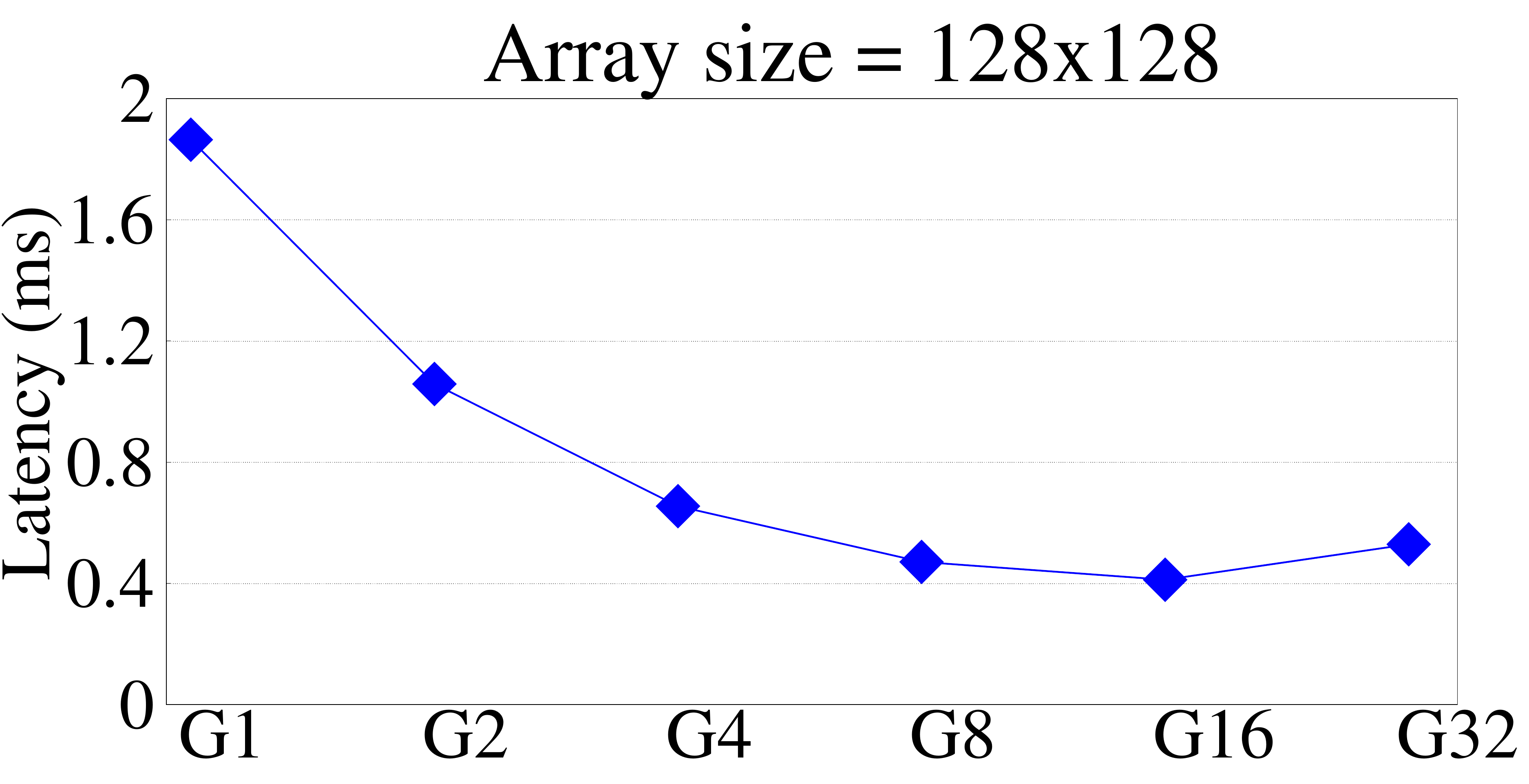} \\
\includegraphics[scale=0.116]{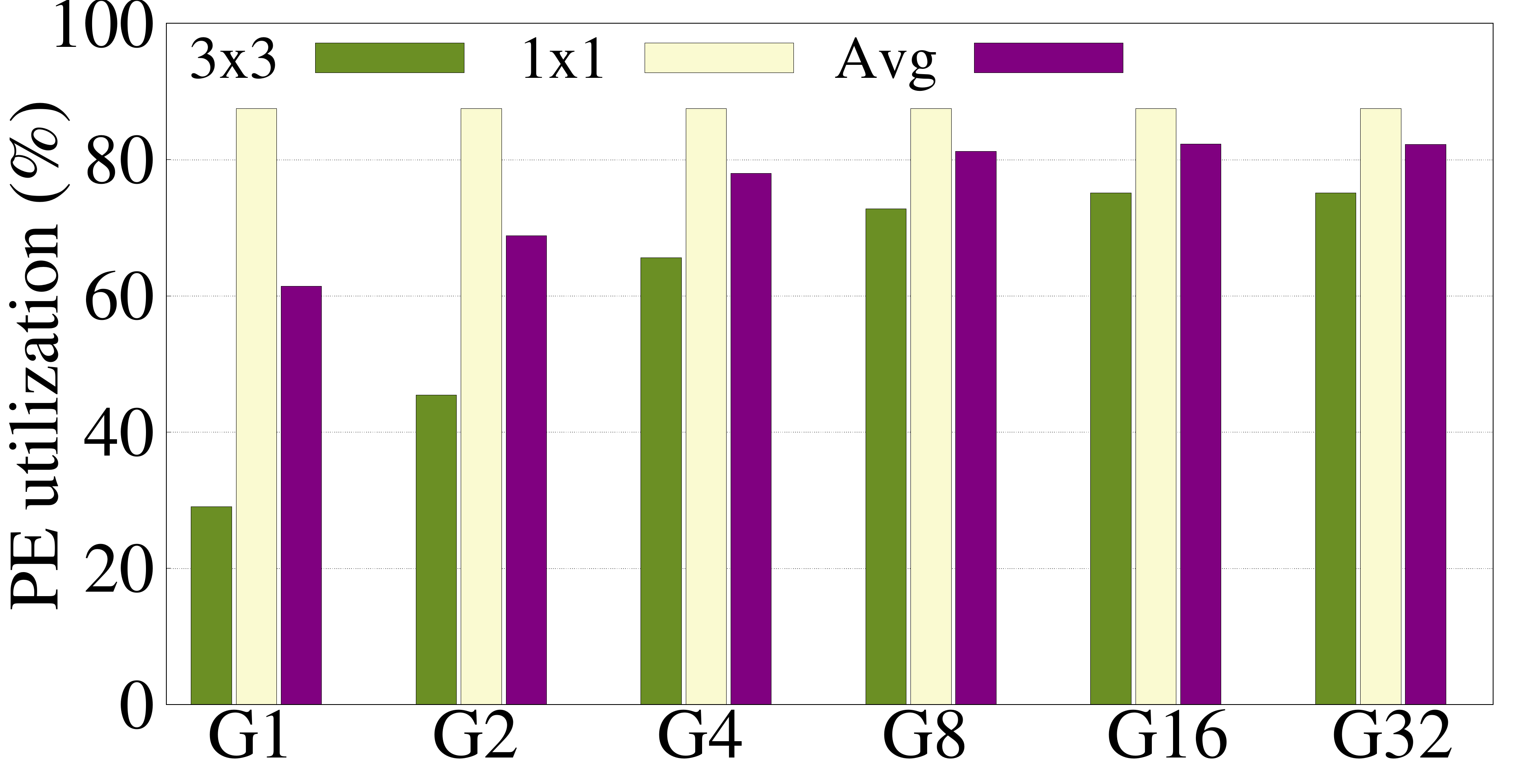}  
\includegraphics[scale=0.116]{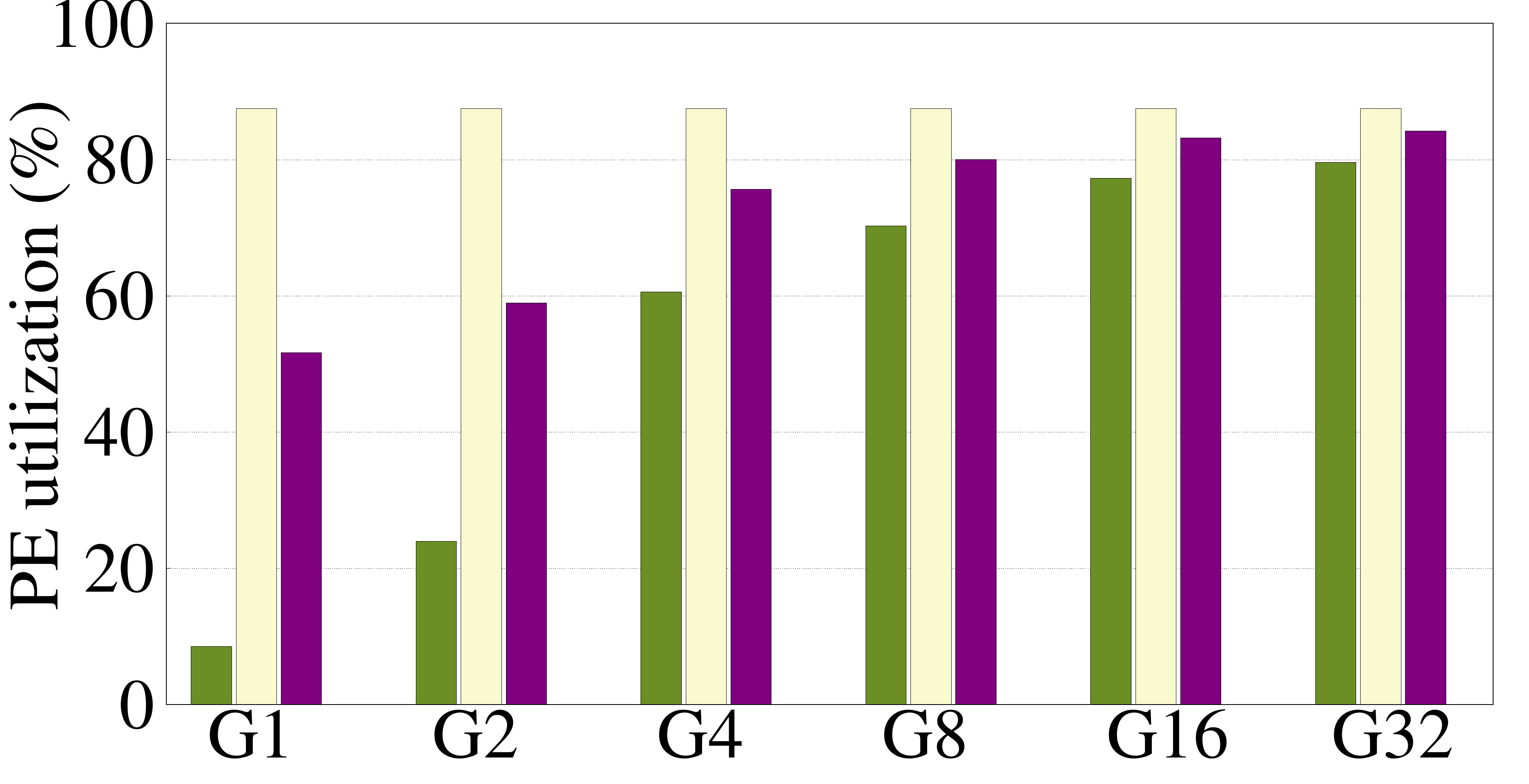} 
\includegraphics[scale=0.116]{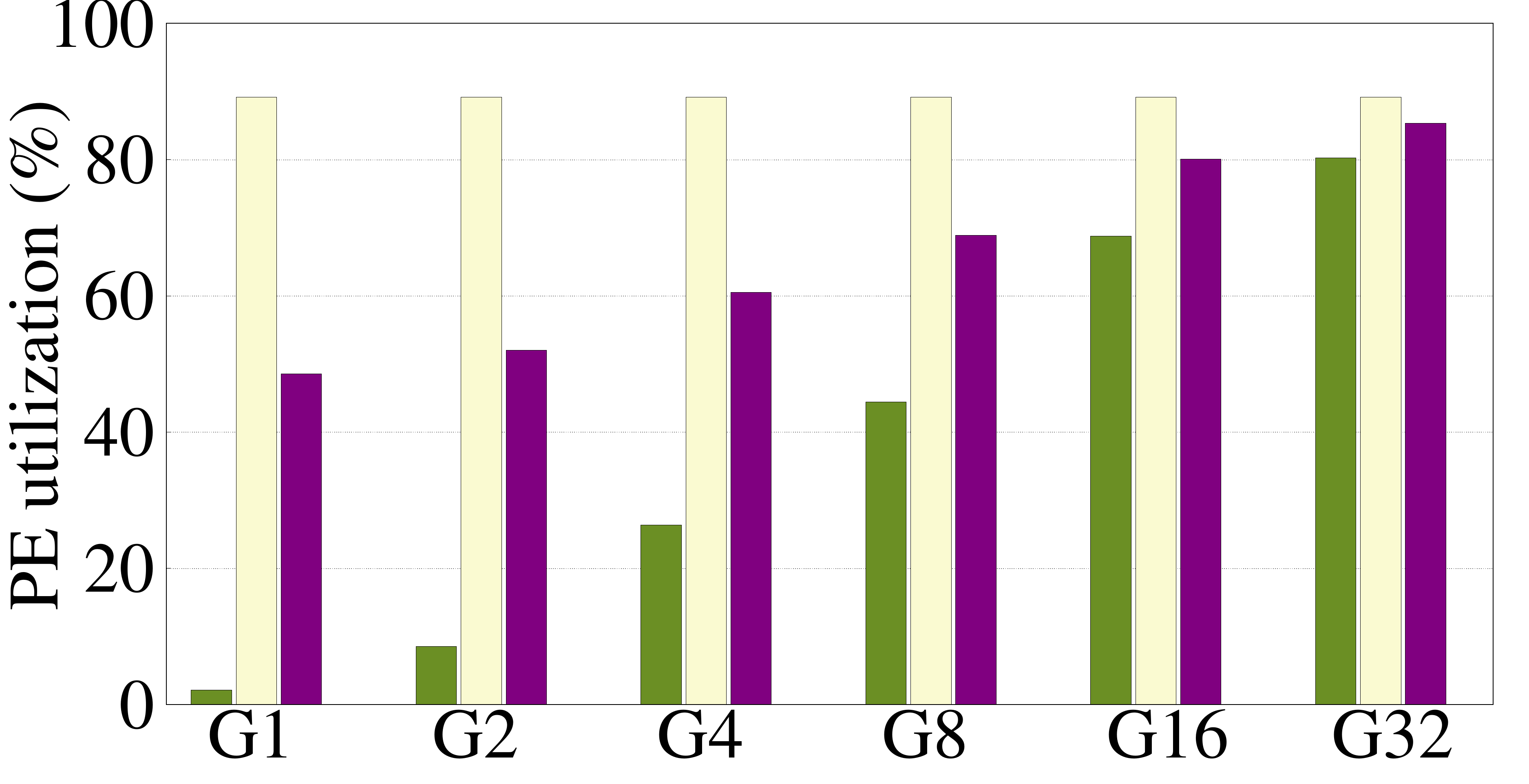}  
\includegraphics[scale=0.116]{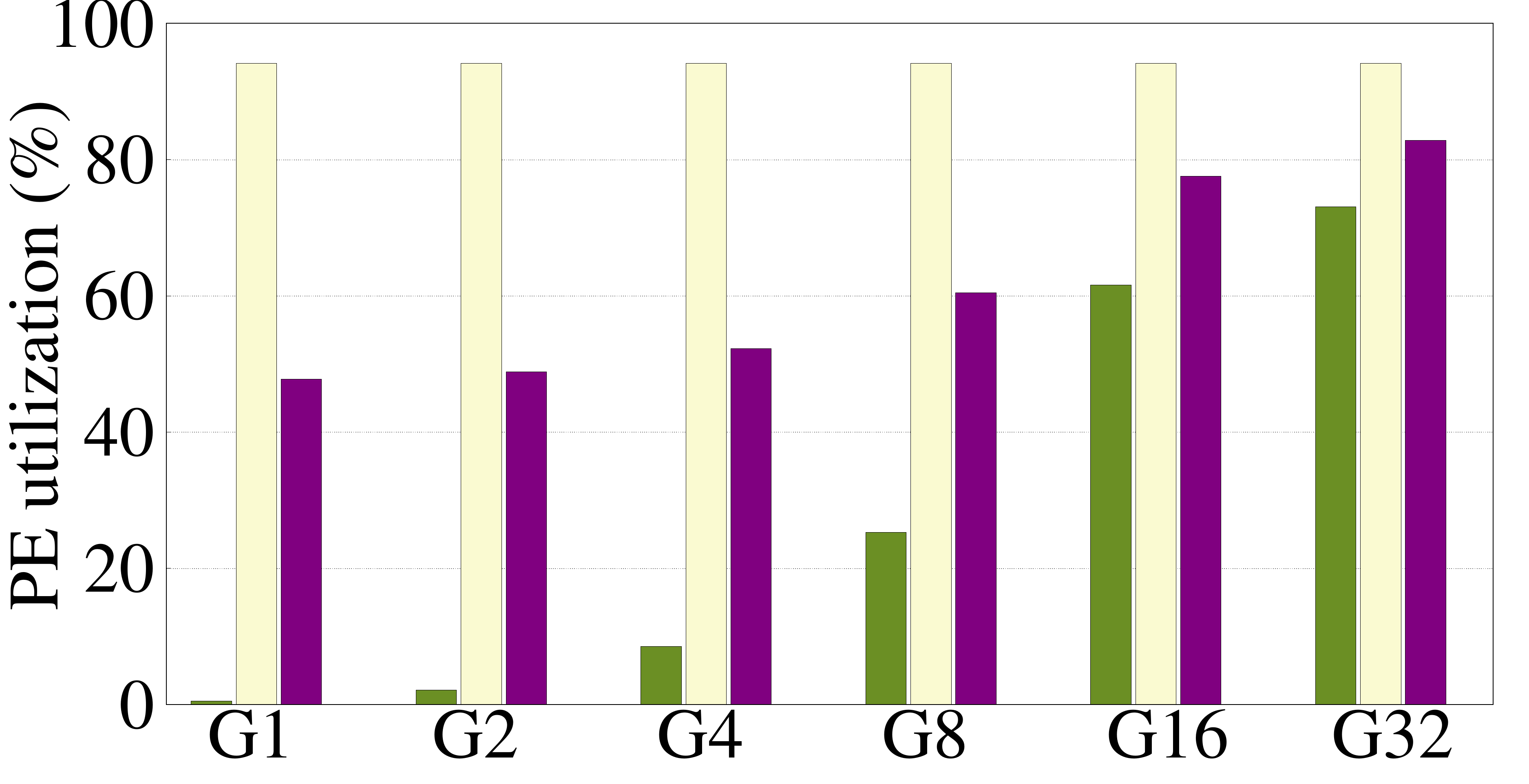}  \\
\includegraphics[scale=0.116]{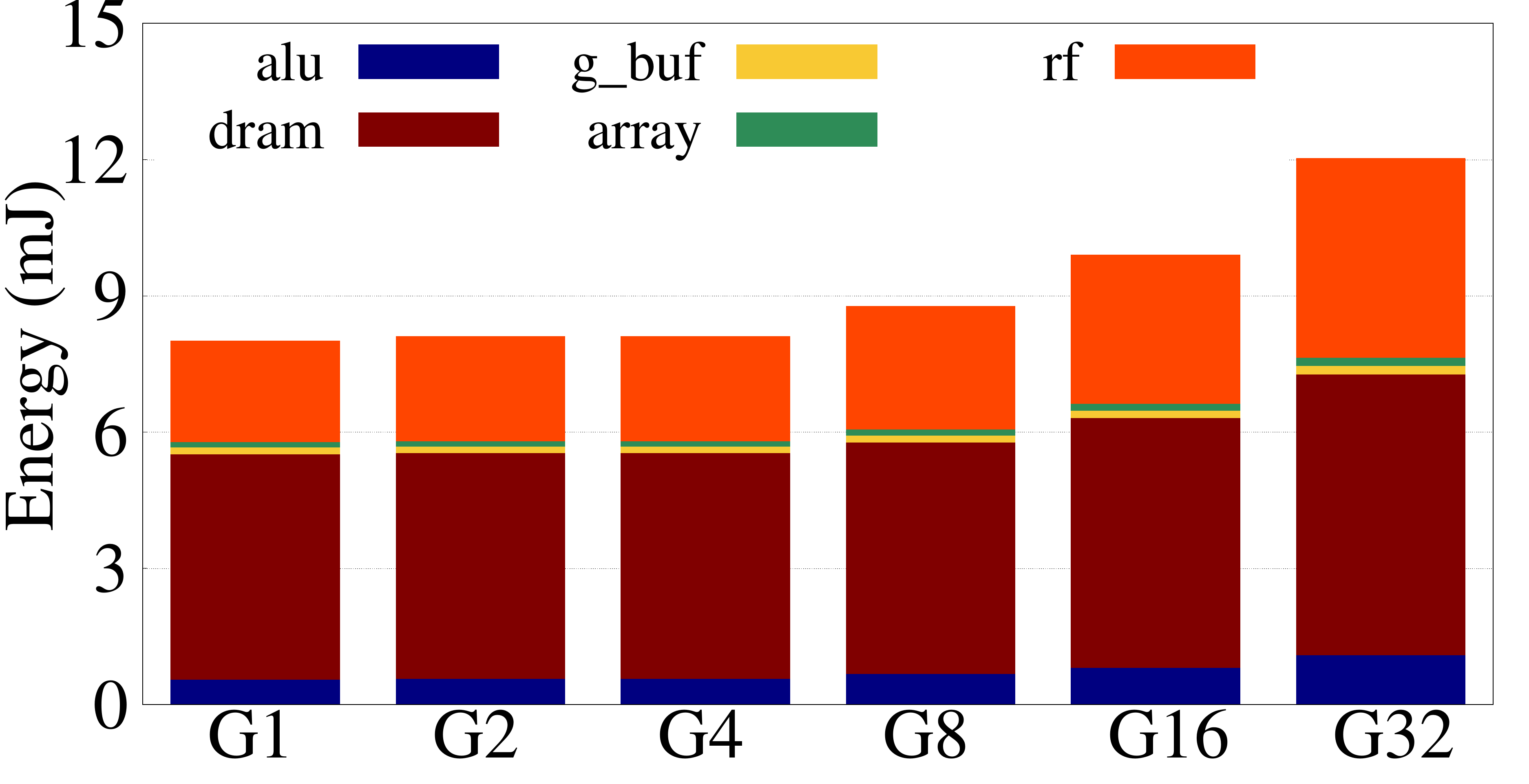}  
\includegraphics[scale=0.116]{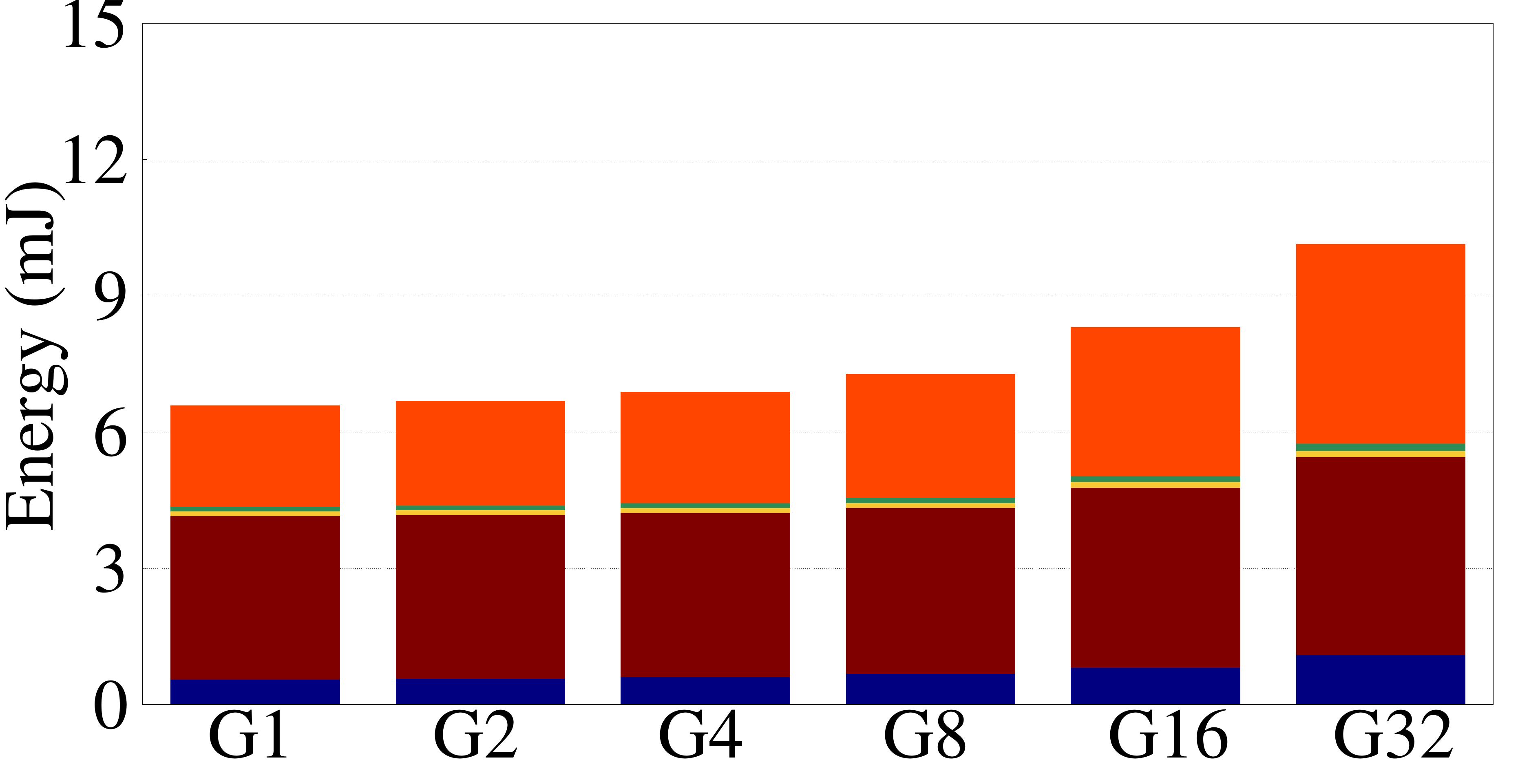} 
\includegraphics[scale=0.116]{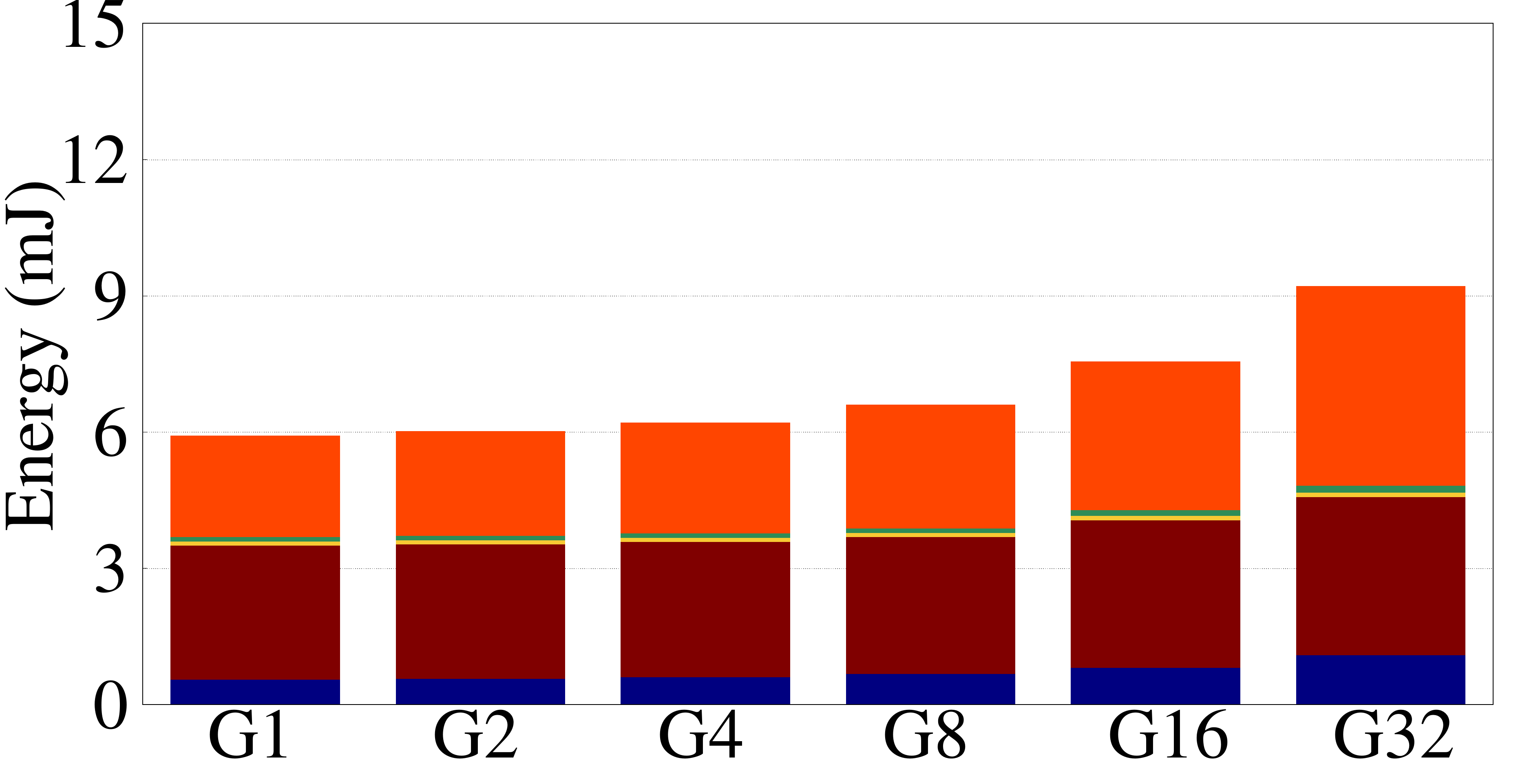}  
\includegraphics[scale=0.116]{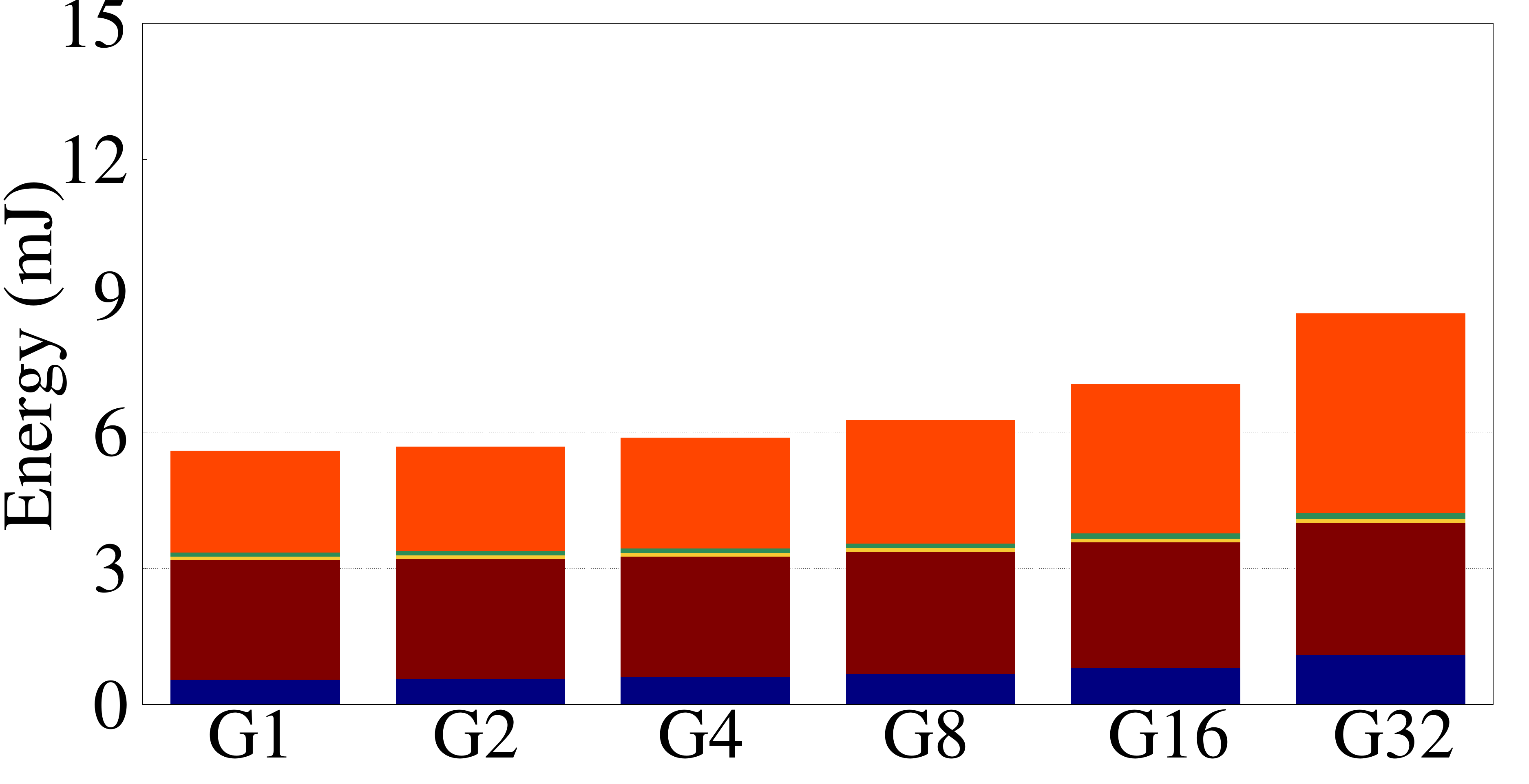} 
\caption{Results for MobileNetV1 ($\alpha$=1, $\rho$=1): In top row latency; in middle row PE utilization for $3\times3$ conv, $1\times1$ conv,  and average utilization; and in bottom row total energy consumption in terms of ALU (alu), DRAM (dram), global buffer (g\_buf), systolic array (array), and register file (rf) accesses energy with variations in $G$ and systolic array size are shown. }
\label{fig:MV1Vanilla}
\end{figure*}



 \begin{table}[htbp]
\caption{MACs (in Millions) and param (\#parameters in Millions) variations in MobileNetV1 with different $\alpha$ and $\rho$ }
\label{tab:MACsParamsMV1All}
\centering
\resizebox{0.49\textwidth}{!}{
\begin{tabular}{|c|l|r|r|r|r|r|r|r|}
\hline
 & {\bf Metric} & {\bf G1} & {\bf G2} & {\bf G4} & {\bf G8} & {\bf G16} & {\bf G32} & {\bf G64}  \\ \hline
\multicolumn{9}{|c|}{(a) Baseline MobileNetV1 ($\alpha$=1, $\rho$=1)} \\
\hline
      &  MACs   & 569   & 586   & 621   & 690   & 830   & 1108  & - \\
\hline
      &  Param   & 4.21  & 4.25  & 4.34  & 4.52  & 4.88  & 5.59  & - \\
\hline
\multicolumn{9}{|c|}{(b) MobileNetV1 with different values of $\alpha$ ($\rho$ set to 1)} \\
\hline
\multirow{2}{*}{$\alpha$=0.5} & MACs  & 147   & 156   & 173   & 208   & 278   & \multicolumn{1}{l|}{ - } & \multicolumn{1}{l|}{ -} \\
\cline{2-9}      &  Param   & 1.82  & 1.83  & 1.85  & 1.9   & 1.99  & \multicolumn{1}{l|}{ - } & \multicolumn{1}{l|}{ -} \\
\hline
\multirow{2}{*}{$\alpha$=2} &  MACs  & 2237  & 2272  & 2341  & 2481  & 2759  & 3315  & 4428 \\
\cline{2-9}      &  Param   & 13.76 & 13.94 & 14.3  & 15.01 & 16.44 & 19.3  & 25.01 \\
\hline
\multicolumn{9}{|c|}{(c) MobileNetV1 with different values of $\rho$ ($\alpha$ set to 1)} \\
\hline
\multirow{2}{*}{$\rho$=0.5} &  MACs   & 143   & 147   & 156   & 173   & 208   & 278   & - \\
\cline{2-9}      &  Param   & 4.21  & 4.25  & 4.34  & 4.52  & 4.88  & 5.59  & - \\
\hline
\multirow{2}{*}{$\rho$=2} &  MACs   & 2272  & 2341  & 2481  & 2759  & 3315  & 4428  & - \\
\cline{2-9}      &  Params   & 4.21  & 4.25  & 4.34  & 4.52  & 4.88  & 5.59  & - \\
\hline
\end{tabular}%
}
\end{table}

We measure the predictive performance of DNNs as the prediction accuracy on the image classification task. We train our models from scratch (i.e. no pre-training), without any data augmentation techniques, on  Imagenette dataset \cite{Fastai_Imagenette}  with input crop size $224\times224$ for 150 epochs. Imagenette is a subset of 10 classes from the popular benchmark dataset ImageNet. Note that all the reported top-1 accuracy are average of three runs to avoid the effect on noise on accuracy.

\subsection{Results}

{\bf PE utilization and latency} In middle row of Fig. \ref{fig:MV1Vanilla}, we show utilization for 1x1 and 3x3 convolutions separately. The reported ``average'' PE utilization is arithmetic mean of that of the all conv, FC, pooling layers in MobileNetV1. At $G$=1, the PE utilization of MobileNetV1 on the smaller array is well above 50\%; however, it reduces on scaling up the array size.  For example, the average PE utilization of MobileNetV1 on $16\times16$ array is 61.4\% while that on $128\times128$ array is 47.8\%. Across all the array sizes, the PE utilization for $1\times1$  conv is well above 80\%, however, it is substantially low for $3\times3$ conv and further decreases with higher array size (e.g., 29\% on $16\times16$ array, and only 2.1\% on $64\times64$ array). Increasing the number of channels in each group (i.e., $G$) of $3\times3$ DWConv improves PE utilization. The overall utilization  reaches  $\approx$80\% on $16\times16$ and $32\times32$ arrays at $G$=8. By contrast, on $64\times64$ and $128\times128$ arrays, utilization reaches  $\approx$80\% at $G$=16.

As shown in Table \ref{tab:MACsParamsMV1All}(a), increasing $G$ increases the number of computations  and hence, the latency on $16\times16$ array increases with higher $G$. However, on larger array sizes, the latency first decreases and reaches minima and further increases with an increase in $G$. On larger array sizes, PE underutilization becomes significant, and utilization improves with an increase in $G$. Hence, the latency starts decreasing initially and reaches a minimum value; however, it further increases as the number of MACs becomes significantly higher. Clearly, at lower $G$, PE underutilization is significant, and once the utilization is sufficiently high ($\approx$80\%), the latency is driven by the computational complexity of DNN. Notice that with the increase in array sizes, the minima in latency curve shifts towards the right, i.e., towards higher $G$. This happens because a higher number of channels is required to utilize the higher number of PEs on a larger array. Hence, the latency is driven by PE utilization, and better PE utilization overwhelms the effect of the increase in computational complexity, up to a certain value of $G$.

{\bf Key takeaway 1:} {\em The optimum  latency is driven by both the  PE utilization as well as the computational complexity of DNN, and the effect of PE utilization on latency depends on PE array size in the systolic accelerator.}

We plot the total energy consumption with different $G$ values to see the energy overhead of increasing $G$ (bottom row in Fig. \ref{fig:MV1Vanilla}). Since increasing $G$ leads to a gradual increase in the number of computations, there is a negligible increase in the inference energy up to a group size of $G$=8. Moreover, increasing $G$ causes better data locality, and hence, the increase in computation results only in more number of access to RF rather than DRAM. Thus, even at a higher $G$, the overall increase in energy is not substantial.   

\begin{figure}[htbp]
\centering
\includegraphics[scale=0.112]{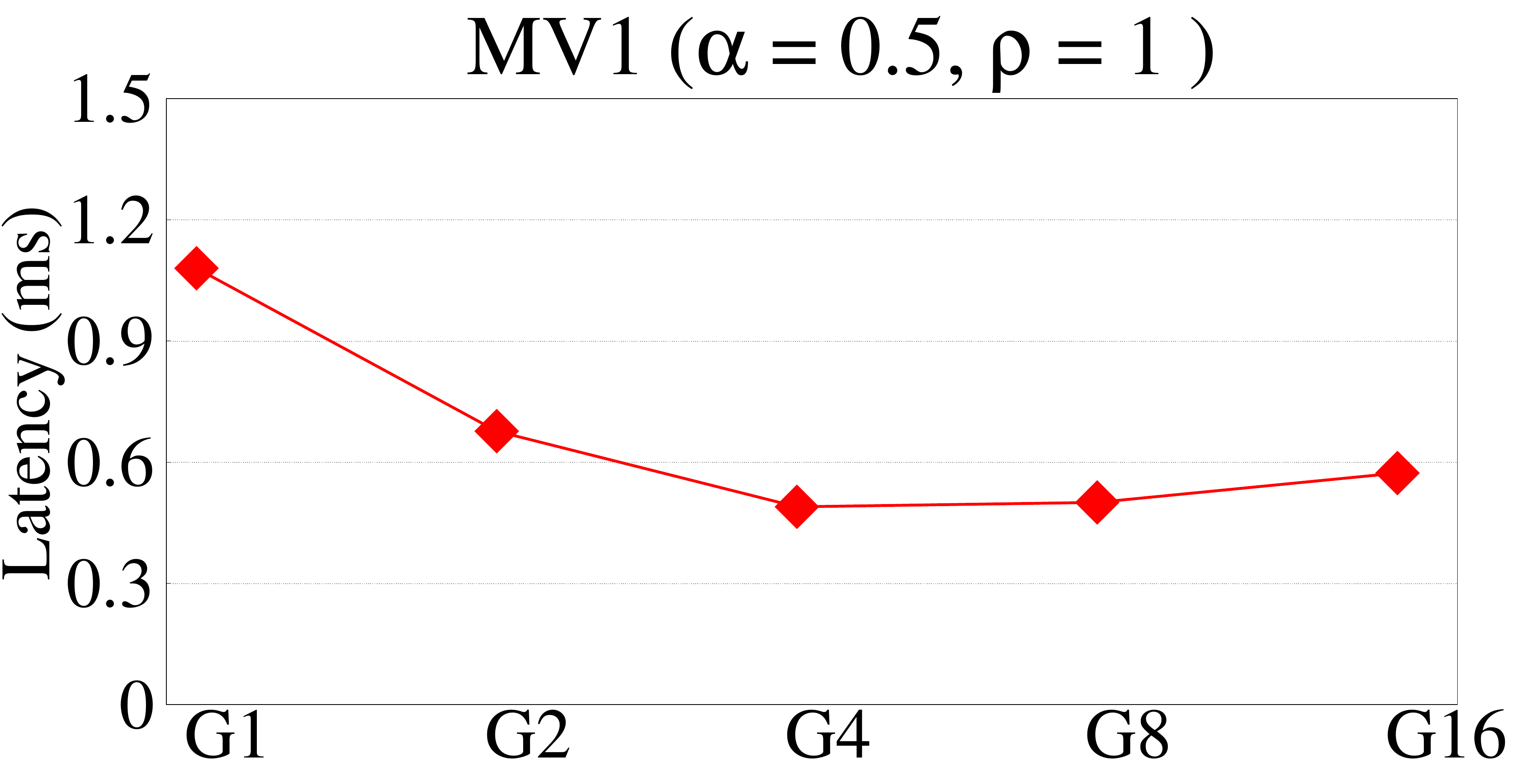}  
\includegraphics[scale=0.112]{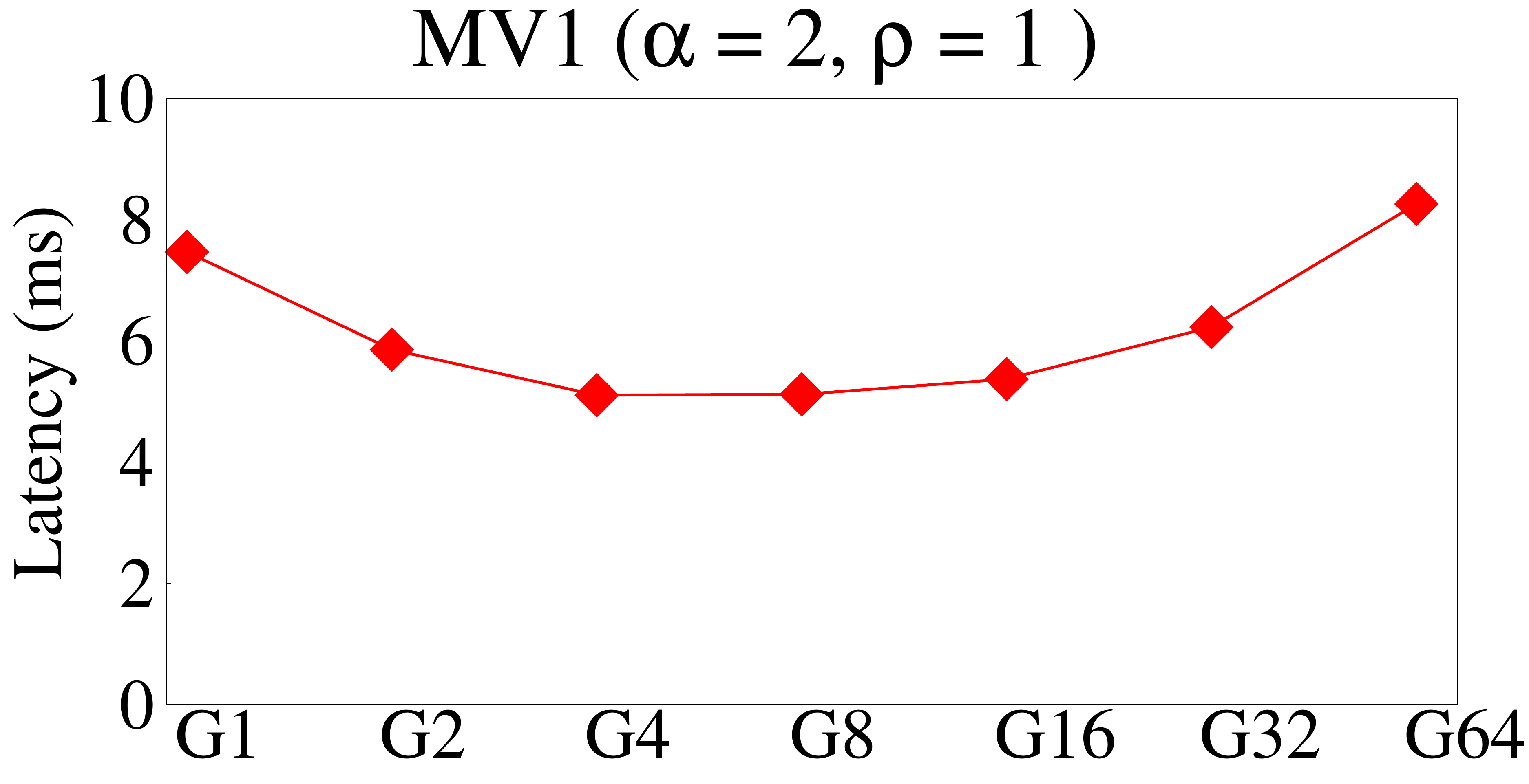} \\
\includegraphics[scale=0.112]{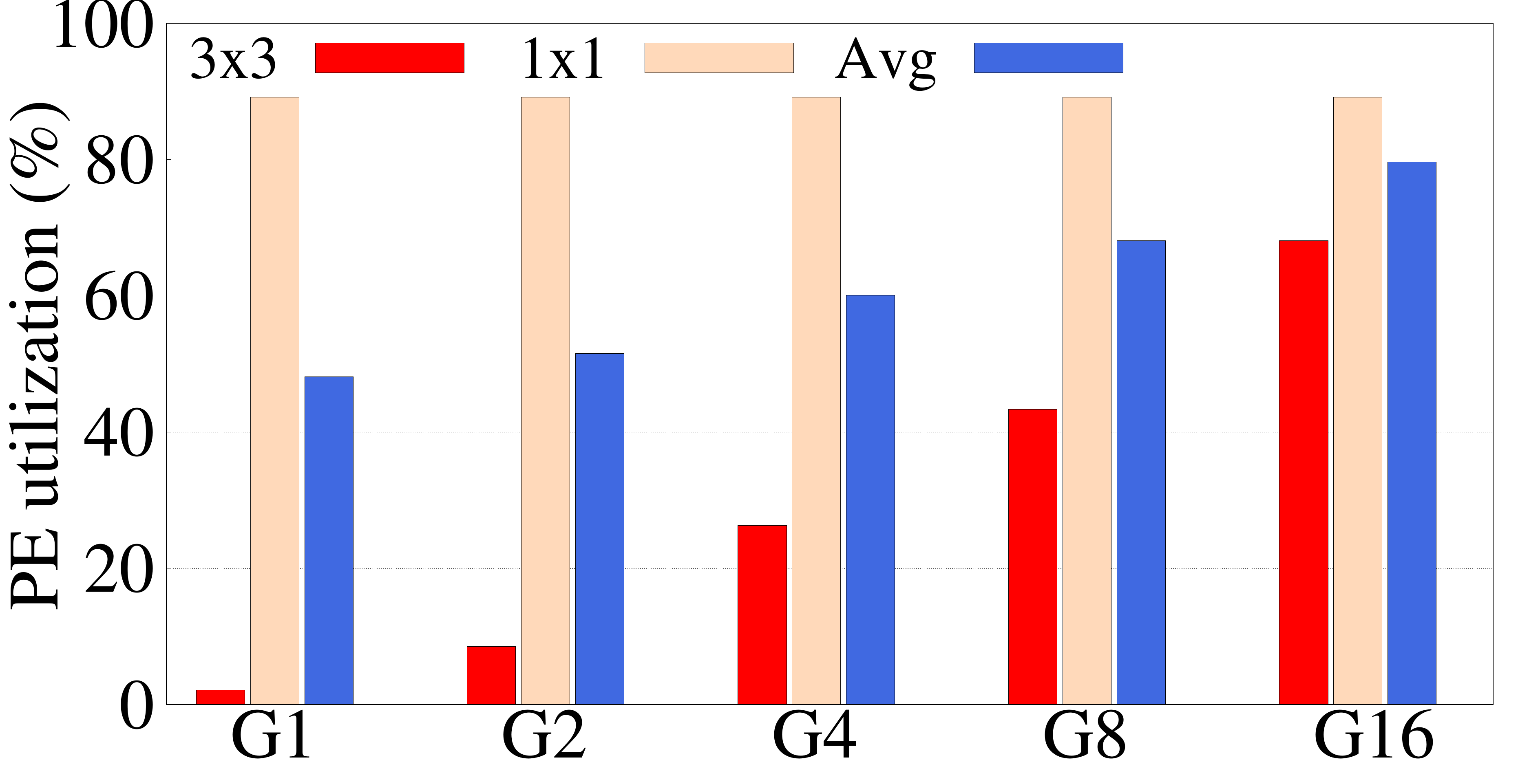}  
\includegraphics[scale=0.112]{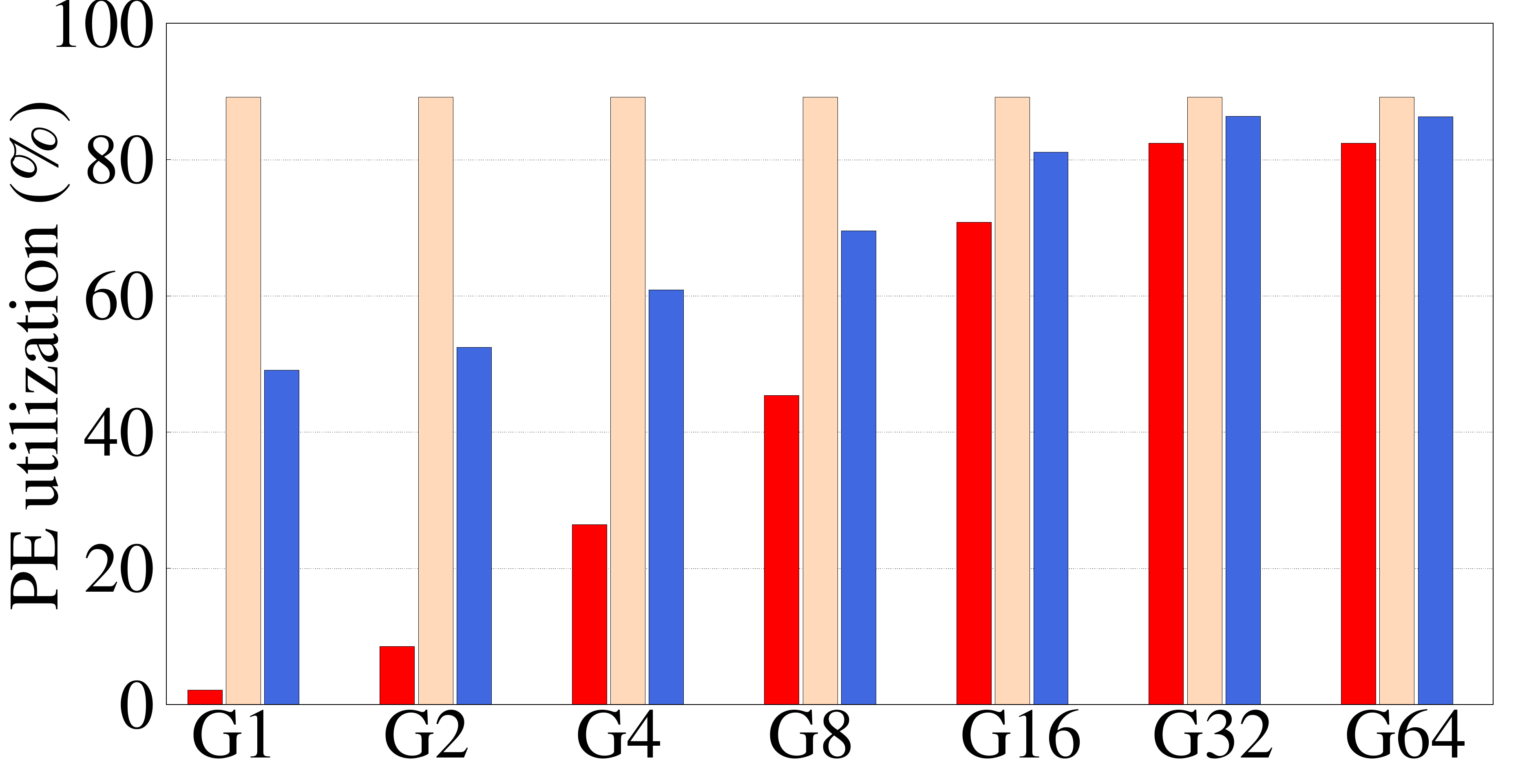} \\
\includegraphics[scale=0.112]{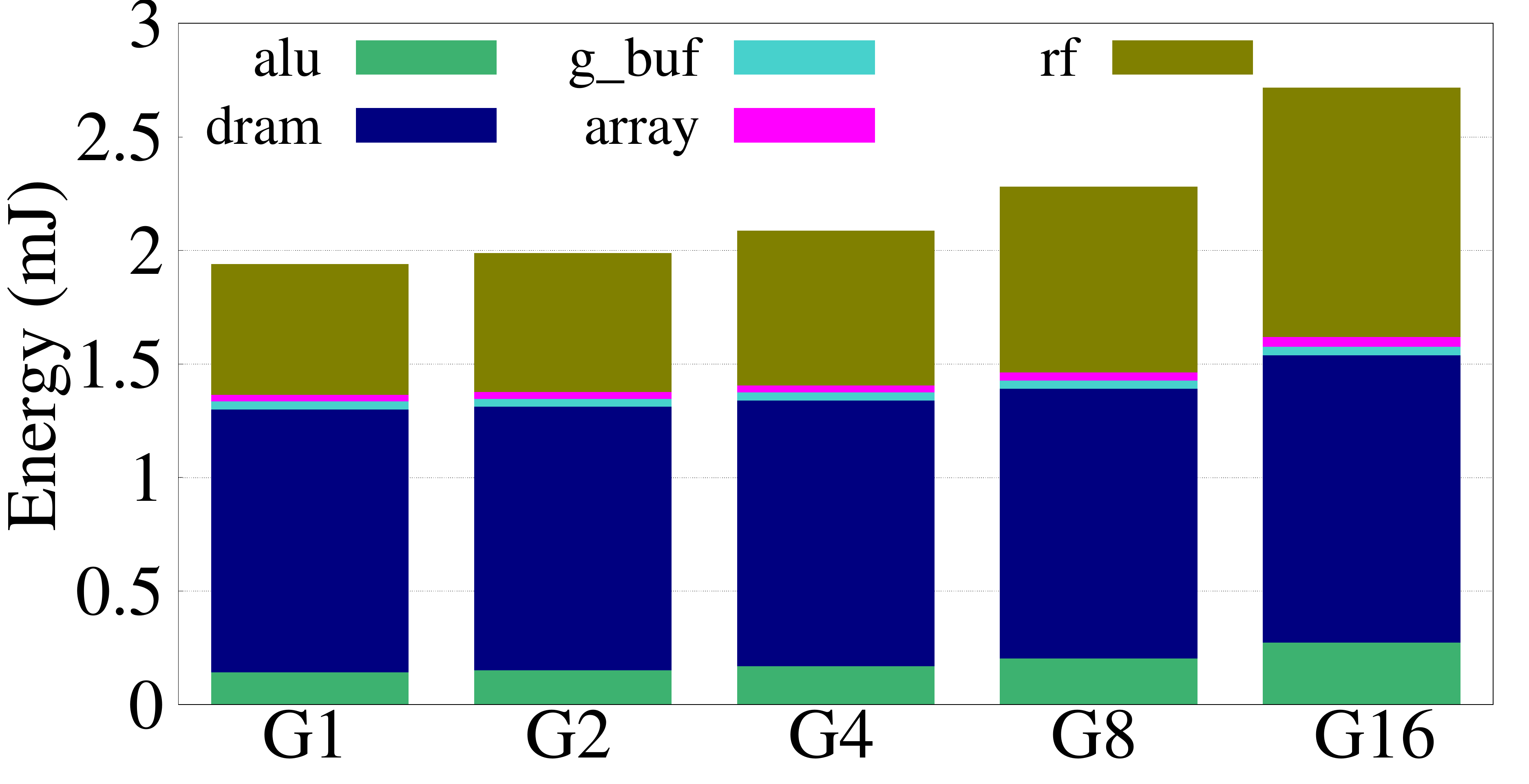}  
\includegraphics[scale=0.112]{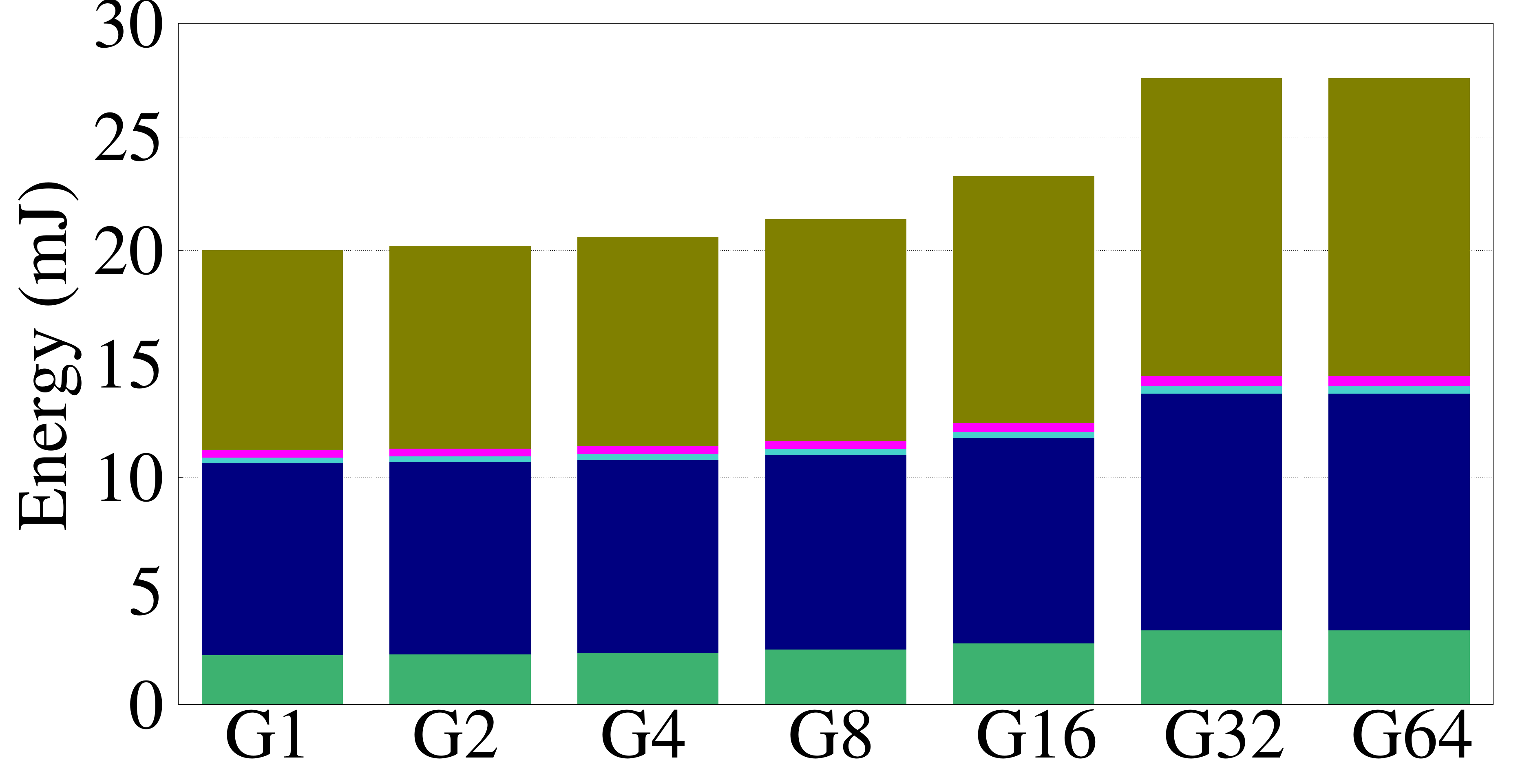} 
\caption{Results for MobileNetV1 with $\alpha$ = 0.5 (left column), and $\alpha$ = 2 (right column) on $64\times64$  PE array: In top row latency; in middle row PE utilization; and in bottom row total energy consumption variations with different $G$ are shown. }
\label{fig:MV1Alpha}
\end{figure}

{\bf Effects of changing the number of filters} We set $\alpha$ = 0.5 to halve the number of filters in all the layers and set $\alpha$ = 2 to double the number of filters.
Since changing $\alpha$ does not change the number of channels per group, the PE utilization remains the same across different values of $\alpha$ (middle row in Fig. \ref{fig:MV1Alpha}). For $\alpha$=0.5, latency decreases sharply at initial values of $G$ and remains constant at higher $G$. However, for $\alpha$=2, latency decreases gradually and starts increasing at higher $G$. Since the number of MACs in MobileNetV1 with $\alpha$=0.5 is quite low (Table \ref{tab:MACsParamsMV1All}(b)), the latency is  driven only by PE utilization and it remains constant at higher $G$. However, at higher $G$   with $\alpha$=2, 
the number of MACs is very high, which results in increased latency. For all values of $G$, the absolute values of latency and energy are significantly higher at $\alpha$=2 compared to those at $\alpha$=0.5. This is because the number of MACs and parameters are much higher for $\alpha$=2 (Table \ref{tab:MACsParamsMV1All}(b)).

{\bf Key takeaway 2:} {\em The extent to which PE utilization affects the latency also depends on the total number of MACs in DNN. In a DNN with very few MACs, the latency  depends only on PE utilization.}

\begin{figure}[htbp]
\centering
\includegraphics[scale=0.112]{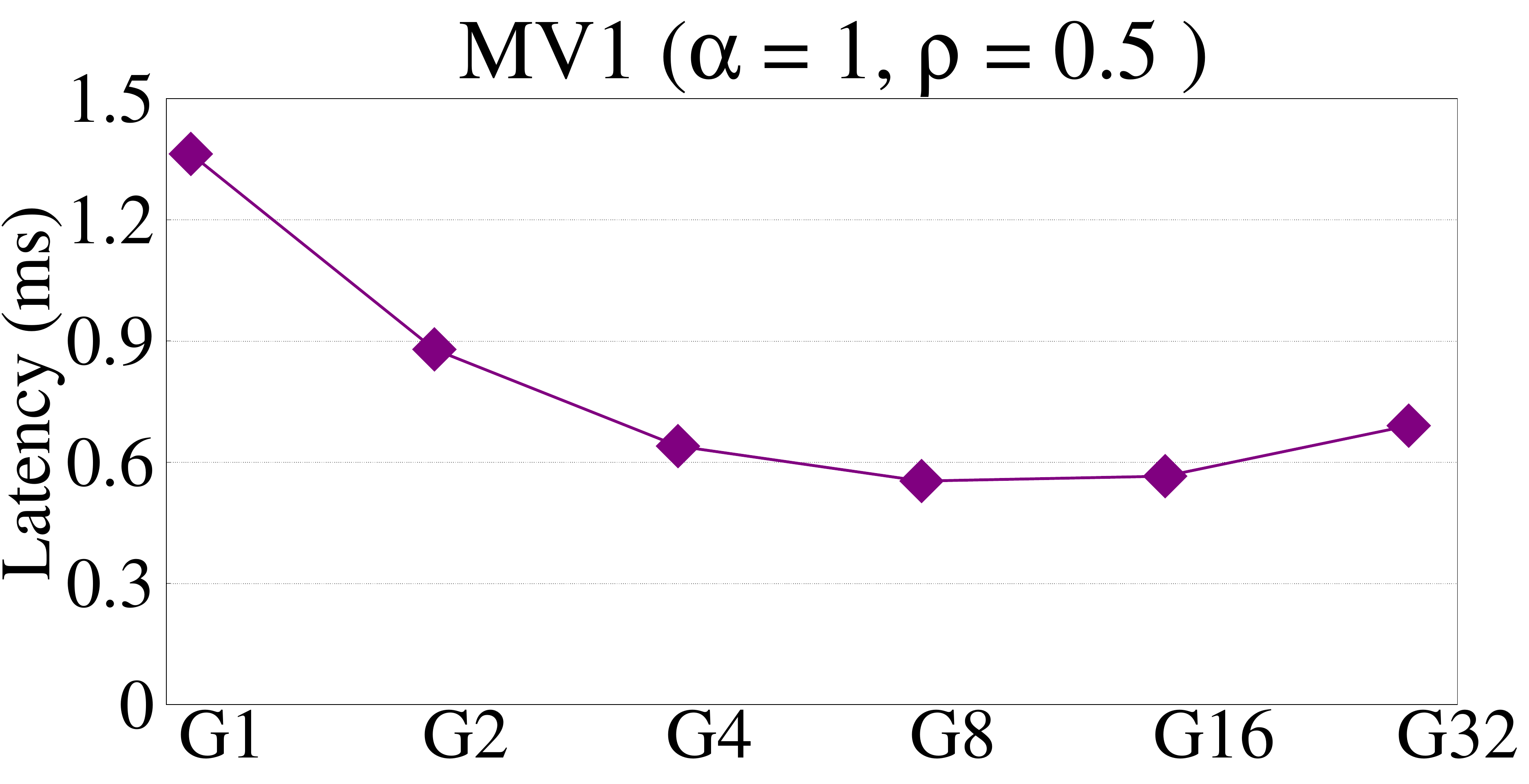}  
\includegraphics[scale=0.112]{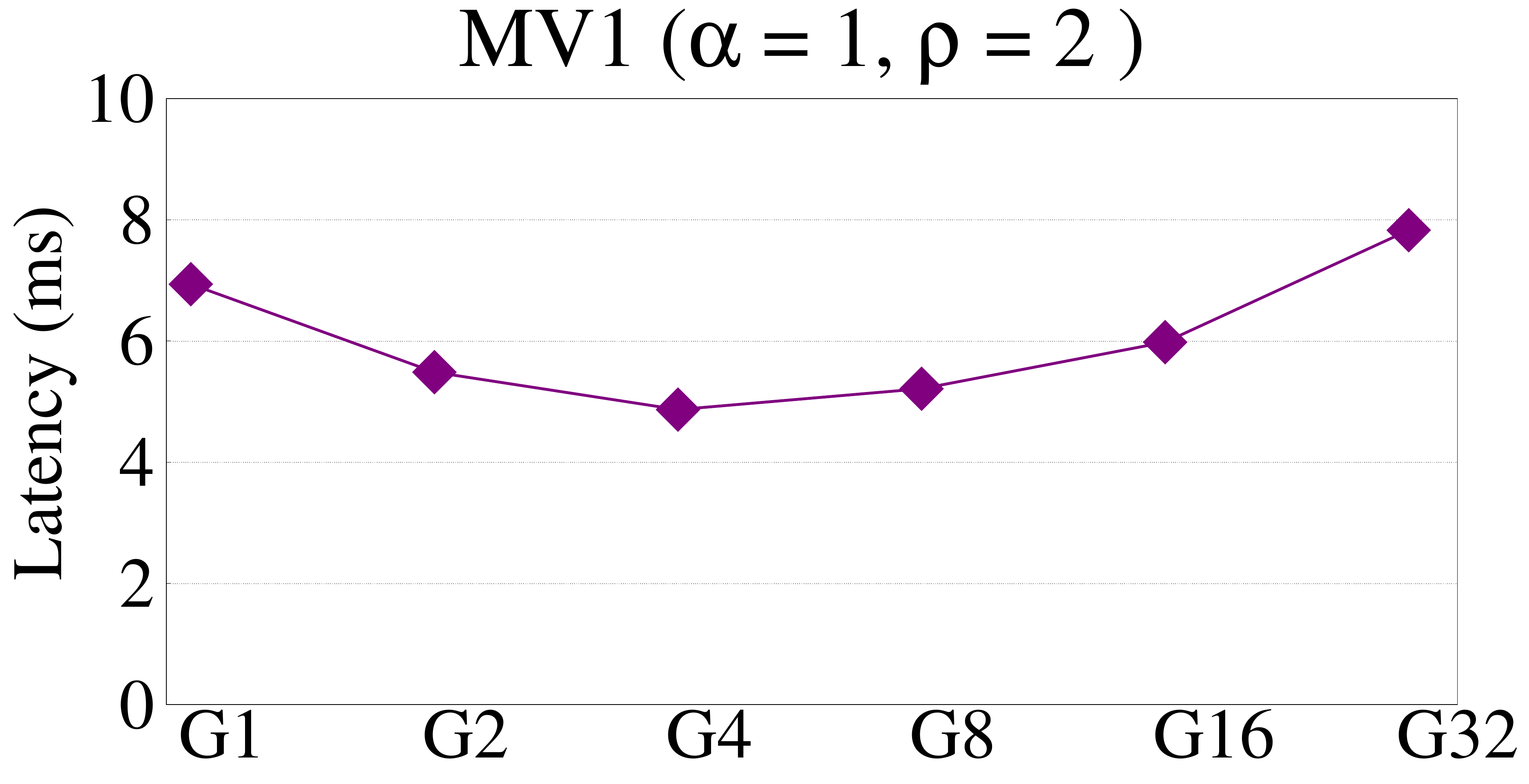} \\
\includegraphics[scale=0.112]{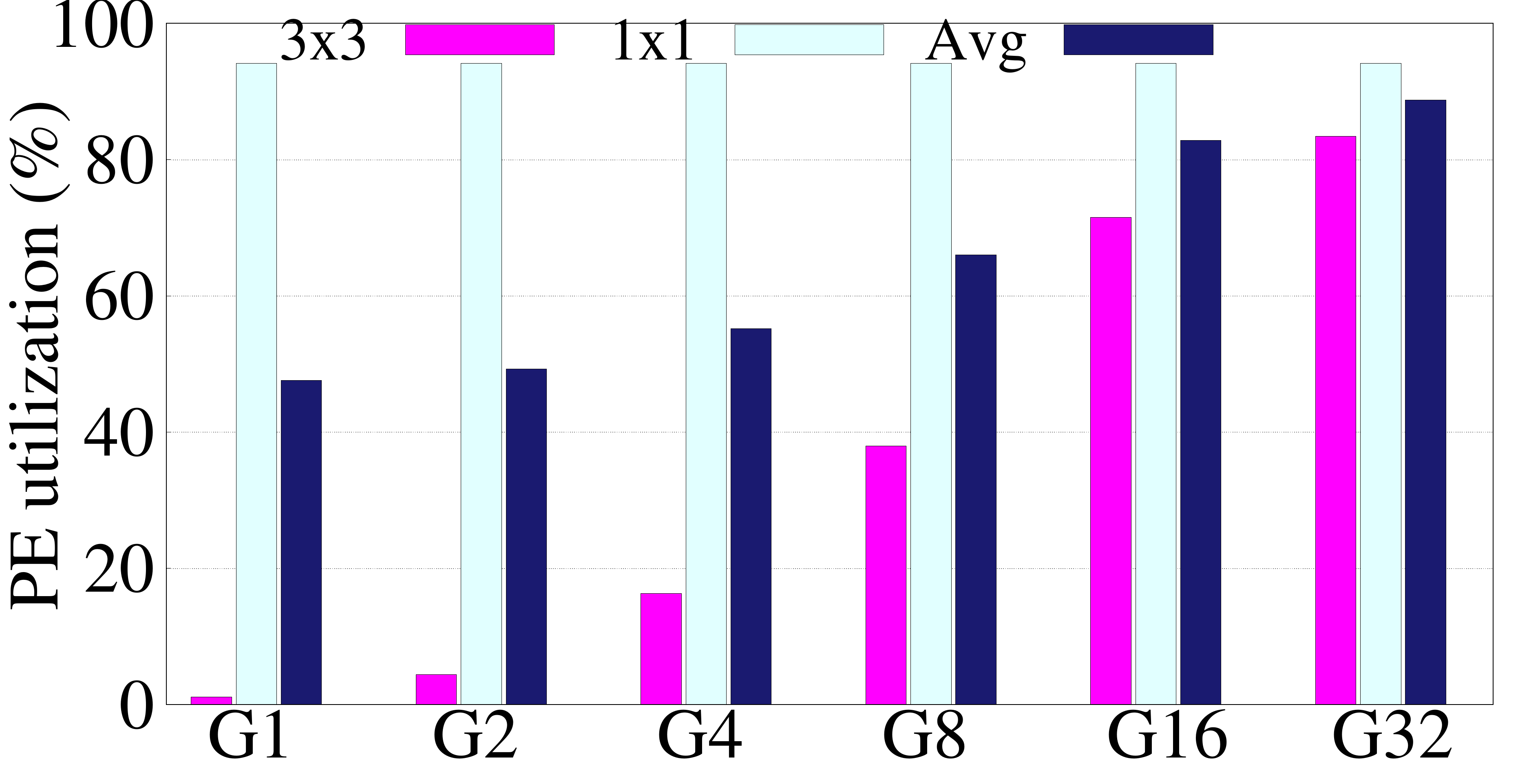}  
\includegraphics[scale=0.112]{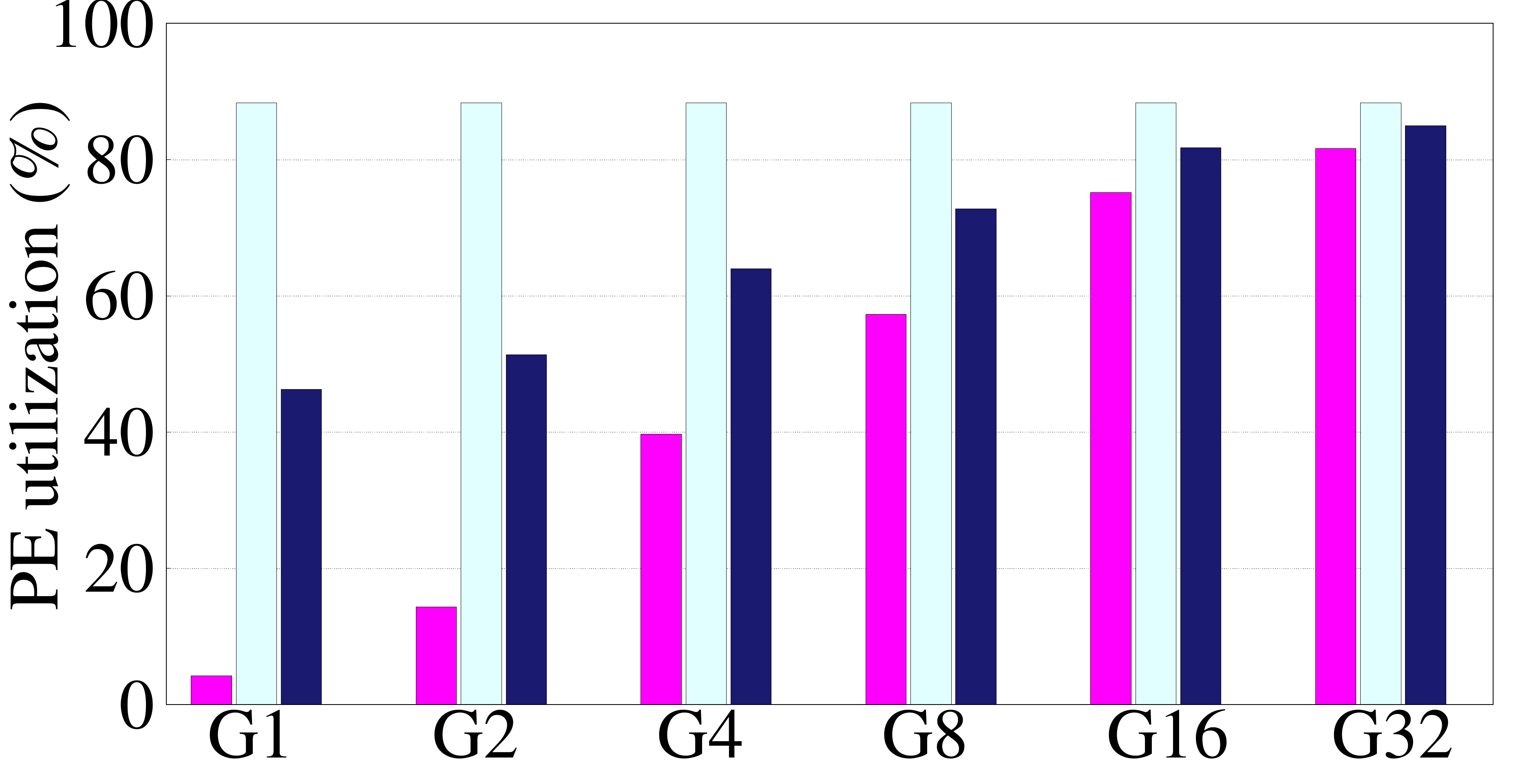} \\
\includegraphics[scale=0.112]{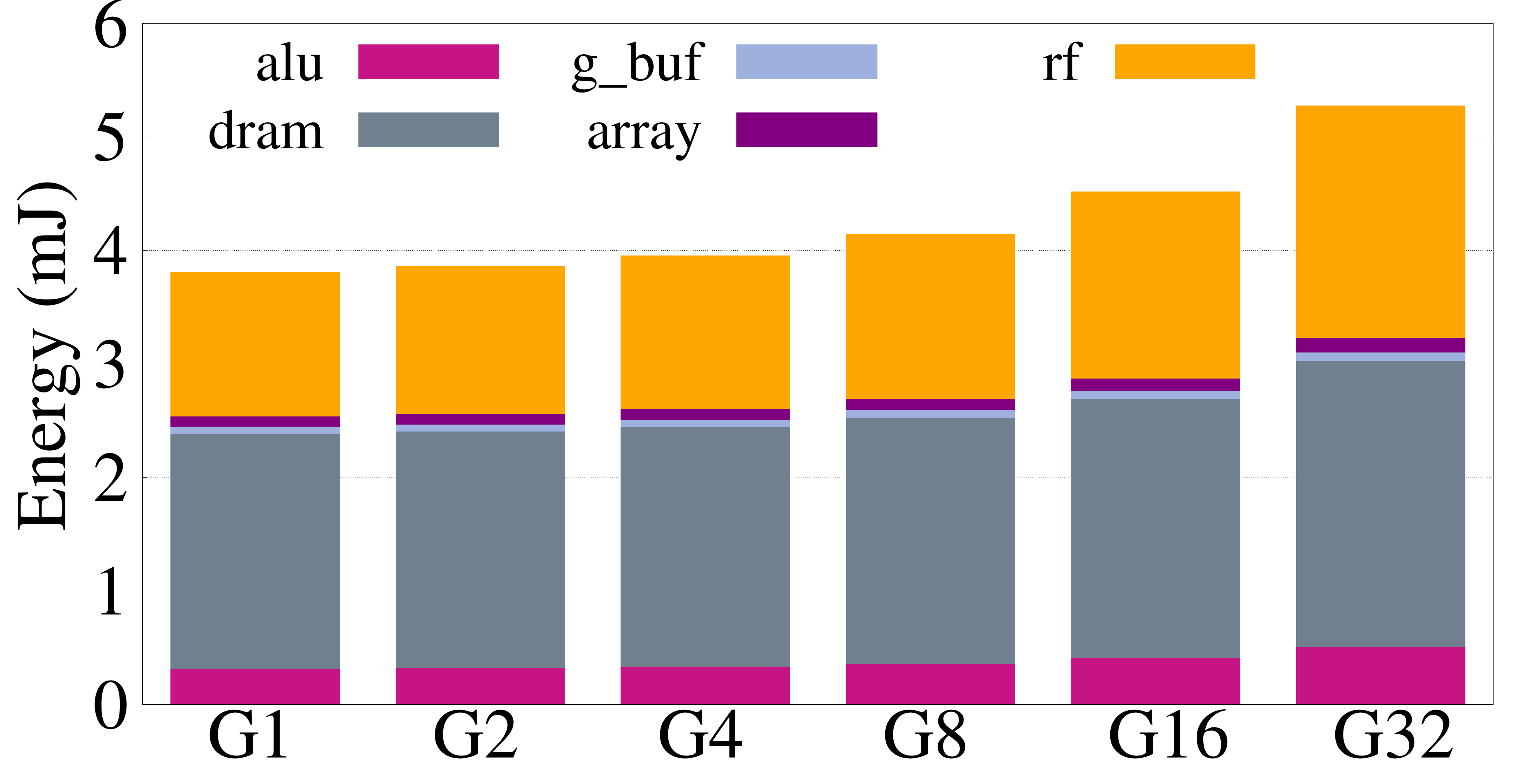}  
\includegraphics[scale=0.112]{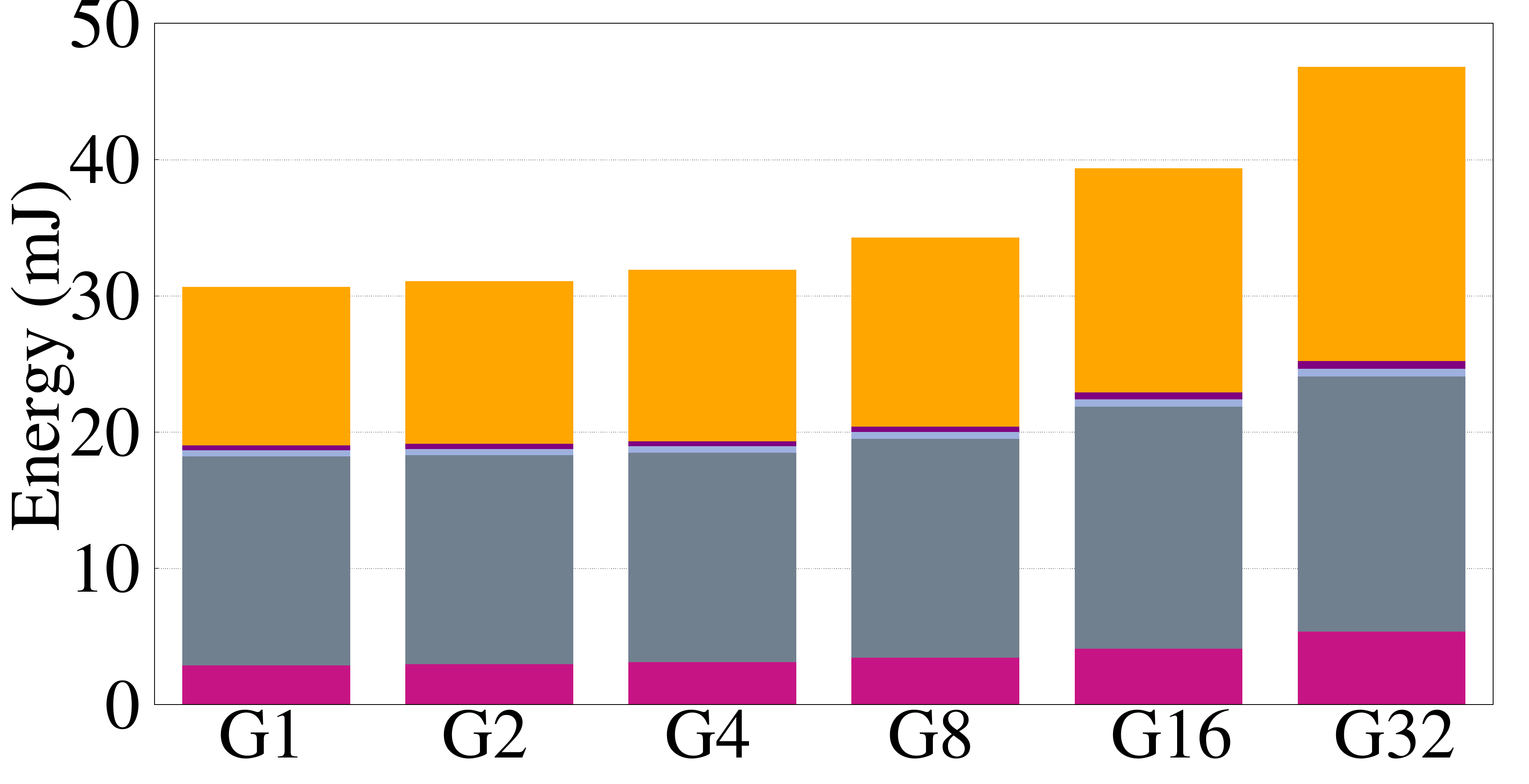} 
\caption{Results for MobileNetV1 with $\rho$ = 0.5 (left column), and $\rho$ =2 (right column) on $64\times64$  PE array: In top row latency; in middle row PE utilization; and in bottom row total energy consumption variations with different $G$ are shown.}
\label{fig:MV1Rho}
\end{figure}

{\bf Effects of changing input resolution} We change $\rho$ to 0.5 and 2  to increase and decrease (respectively) the spatial size of the input image. Changing the input resolution changes the spatial size of fmaps in all the layers, which, in turn, affects the PE utilization. As shown in Fig. \ref{fig:MV1Rho}, due to the higher spatial size of fmaps at $\rho$=2,  PE utilization is higher than that at $\rho$=0.5.   With $\rho$=0.5, the latency decreases sharply at lower $G$, then reaches the minimum value at $G$=8 and $G$=16, and then starts increasing at higher $G$. However, with $\rho$=2, latency decreases gradually at lower $G$, then reaches a minimum at $G$=4, and then starts increasing sharply at higher $G$. In other words, with an increasing value of $\rho$ (from 0.5 to 2), the minima in the latency curve shift towards left (i.e., towards lower $G$).

Thus, when utilization is improved by employing fmap with larger spatial size in a DNN ($\rho$=2), then  benefit of better PE utilization is overshadowed by a substantial increase in computation (Table \ref{tab:MACsParamsMV1All}(c)). For instance, even when a gain in PE utilization from $G$=4 to $G$=16 is $\approx$27\%, the latency is increased by $\approx$23\%. However, with $\rho$=0.5, the number of MACs is quite low, and the effect of PE utilization on latency is noticeable. The absolute latency and energy values with $\rho$=2 are substantially higher than those with $\rho$=0.5 due to the higher computational complexity in the former. 

{\bf Key takeaway 3:} {\em Increasing PE utilization at the expense of a substantial increase in the number of computations does not lead to lower latency, and the effect of higher PE utilization is dominated by the number of computations.}

\subsection{Proposed Alternative for Latency Optimization} \label{sec:BetterAlternative}
 
So far, we have seen that increasing PE utilization by employing more number of channels in each group of a DNN decreases the latency at the cost of a gradual increase in computations. However, increasing PE utilization at the expense of a substantial increase in computation (e.g., MobileNetV1 with $\rho$=2) does not effectively reduce the latency of a DNN. Now, we propose a better alternative for increasing PE utilization, which also reduces the latency more effectively.

{\bf MobileNetV1 with $\alpha$=0.5 and $\rho$=2}  The PE utilization depends on the number of channels and the spatial size of fmaps in a group. Hence,  scaling the number of filters (using $\alpha$) in the layers of a DNN changes only the computational complexity and does not alter the PE utilization. We propose to use $\alpha$ to reduce the computational complexity of MobileNetV1 with higher $\rho$. The appropriate value of $\alpha$ depends on the additional gain in computational complexity, i.e., in a proportion of $\rho$. Here, we set $\alpha$=0.5 to offset the gain in computational complexity in MobileNetV1 with $\rho$=2 without hampering the PE utilization. This reduces the  number of MACs by $\approx$4$\times$ and the number of parameter by a factor of $\approx$2.5$\times$ (Table \ref{tab:MACsParamsMV1All}(c) and Table \ref{tab:MV10.5Alpha2Rho}). The substantial reduction in both the number of MACs and parameters leads to an improvement in both energy efficiency and latency for all the values of $G$ (Table \ref{tab:Latency&EnergyComp}). This improvement is higher on smaller array sizes because the effect of computational complexity on latency is significant when the array size is smaller. However, the effect of PE utilization on latency is significant on a larger array.  

\begin{table} [htbp] 
\caption{Number of MACs and parameters in MobileNetV1 with $\alpha$ = 0.5, and $\rho$ = 2 } 
\label{tab:MV10.5Alpha2Rho}\centering
\begin{tabular}{|c|c|c|c|c|c|} \hline
{\bf Metric} & {\bf G1} & {\bf G2} & {\bf G4} & {\bf G8} & {\bf G16} \\ \hline
\#MACs (M) & 586 & 621 & 690 & 830 & 1108 \\ \hline
\#Params (M) & 1.82 & 1.83 & 1.85 & 1.90 & 1.99 \\ \hline
\end{tabular} 
\end{table}

{\bf Key takeaway 4:} {\em The computational complexity of a DNN can be reduced without affecting the PE utilization.  At constant PE utilization, the effect of computational complexity on latency is quite significant on smaller array sizes. }

\begin{table} [htbp] 
\caption{Performance comparison of MobileNetV1 versions \{$\alpha$ = 1, and $\rho$ = 2\} and \{$\alpha$ = 0.5, and $\rho$ = 2\}  } 
\label{tab:Latency&EnergyComp}\centering
\resizebox{0.48\textwidth}{!}{
\begin{tabular}{|c|C{1cm}|c|c|c|c|c|c|} \hline
{\bf Model} & {\bf Array size} & {\bf Metric} & {\bf G1} & {\bf G2} & {\bf G4} & {\bf G8} & {\bf G16} \\ \hline
\multirow{12}{*}{\parbox{0.8cm}{ MV1  $\alpha$=1, $\rho$=2 }} & \multirow{3}{*}{ 16x16 } & PE util. (\%) & 68 & 77 & 79 & 79 & 80 \\
& & Latency (ms) & 66.5 & 67.7 & 72.1 & 81.1 & 99.2 \\
& & Energy (mJ) & 59.7 & 60.1 & 61.0 & 63.7 & 69.3 \\ \cline{2-8}
& \multirow{3}{*}{ 32x32 } & PE util. (\%) & 56 & 65 & 77 & 82 & 83 \\ 
& & Latency (ms) & 18.8 & 17.8 & 18.2 & 20.1 & 24.4 \\
& & Energy (mJ) & 37.2 & 37.6 & 38.4 & 41.1 & 46.6 \\ \cline{2-8}
& \multirow{3}{*}{ 64x64 } & PE util. (\%) & 50 & 55 & 66 & 74 & 83 \\
& & Latency (ms) & 6.9 & 5.5 & 4.9 & 5.2 & 6.0 \\
& & Energy (mJ) & 30.6 & 31.1 & 31.9 & 34.3 & 39.3 \\ \cline{2-8}
& \multirow{3}{*}{ 128x128 } & PE util. (\%) & 46 & 48 & 54 & 64 & 77 \\
& & Latency (ms) & 4.0 & 2.5 & 1.7 & 1.5 & 1.6 \\
& & Energy (mJ) & 27.2 & 27.6 & 28.5 & 30.2 & 33.9 \\ \hline
\multirow{12}{*}{\parbox{0.8cm}{ MV1  $\alpha$=0.5, $\rho$=2 }} & \multirow{3}{*}{ 16x16 } & PE util. (\%) & 68 & 76 & 79 & 79 & 80 \\
& & Latency (ms) & 17.8 & 18.3 & 20.5 & 25.1 & 34.1 \\
& & Energy (mJ) & 17.5 & 17.7 & 18.2 & 19.5 & 21.6 \\ \cline{2-8}
& \multirow{3}{*}{ 32x32 } & PE util. (\%) & 55 & 65 & 77 & 82 & 83 \\
& & Latency (ms) & 5.5 & 5.0 & 5.2 & 6.2 & 8.3 \\
& & Energy (mJ) & 11.9 & 12.1 & 12.5 & 13.9 & 15.9 \\ \cline{2-8}
& \multirow{3}{*}{ 64x64 } & PE util. (\%) & 49 & 54 & 66 & 73 & 81 \\
& & Latency (ms) & 2.5 & 1.8 & 1.5 & 1.7 & 2.1 \\
& & Energy (mJ) & 10.3 & 10.5 & 10.9 & 12.1 & 14.1 \\ \cline{2-8}
& \multirow{3}{*}{ 128x128 } & PE util. (\%) & 45 & 47 & 52 & 63 & 76 \\
& & Latency (ms) & 1.8 & 1.0 & 0.6 & 0.5 & 0.5 \\
& & Energy (mJ) & 9.2 & 9.4 & 9.8 & 10.7 & 12.4 \\ \hline
\end{tabular}}
\end{table}

\subsection{Implication on Prediction Accuracy}

We now investigate the ramification of changing $G$ on the predictive performance of DNN. As shown in Table \ref{tab:Accuracy}, the top-1 accuracy of baseline MobileNetV1 ($\alpha$=1, $\rho$=1)  increases for lower values of $G$, reaches a maximum value at $G$=4, then starts decreasing at higher $G$. The DWConv only captures the spatial correlation, and the following $1\times1$ convolution captures the channel correlation \cite{guo2019depthwise}. That is, employing only one channel in a group (in DWConv) captures only one variation of a visual concept in the ofmaps. By contrast, increasing the number of channels in the groups of a filter captures more variations of a single visual concept, which, in turn, boosts the representational power of the network and improves the accuracy \cite{Ioannou_2017_CVPR}. Also, at lower $G$, each ofmap is connected to very few ifmaps and serves as a strong regularizer, which in turn improves the generalization in DNN. Therefore,  once the groups get a sufficient number of channels to capture the variations of the visual concepts, the effect of regularization becomes significant. Hence, at higher $G$, top-1 accuracy starts decreasing due to weaker regularization. 

\begin{table}[htbp]
\caption{ Top-1 accuracy (on Imagenette) for MobileNetV1 with different $\alpha$ and $\rho$}
\label{tab:Accuracy}
\centering
\begin{tabular}{|c|l|c|c|c|c|c|c|c|} \hline
{\bf Models} & {\bf G1} & {\bf G2} & {\bf G4} & {\bf G8} & {\bf G16} & {\bf G32}  \\ \hline
MV1  ($\alpha$=1 $\rho$=1) & 84.08 & 84.55 & 84.65 & 83.46 & 83.40 & 79.94 \\ \hline
MV1  ($\alpha$=1 $\rho$=2) & 84.76 & 84.55 & 84.17 & 84.81 & 83.29 & 82.90 \\ \hline
MV1  ($\alpha$=0.5 $\rho$=2) & 82.61 & 83.54 & 83.70 & 82.71 & 82.29 & - \\  \hline

\end{tabular} 
\end{table}

Note that for larger input size higher $G$ is required to capture more variations of a visual concept. Therefore, top-1 accuracy  in MobileNet-V1 with $\rho$=2 is maximum at $G$=8 (Table \ref{tab:Accuracy}). However, DNNs with very few parameters result in under-fitting at $G=1$, hence top-1 accuracy in  MobileNet-V1 with $\alpha$=0.5, and $\rho$=2 is lower at $G$=1 and starts increasing from $G$=2. The difference between the top-1 accuracy of  MobileNetV1 with \{$\alpha$, $\rho$\} =  \{0.5, 2\}, and  \{1, 2\}
is minimum at $G$=4. Also, the latency is  minimum at same $G$ on a $64\times64$ array (Table \ref{tab:Latency&EnergyComp}). Hence, $G$=4 is a sweet spot for both optimum latency and prediction accuracy. In summary, our proposed alternative for latency optimization in MobileNetV1 with $\rho$=2 does not hamper the accuracy, provided groups in DNN have sufficient $G$ to capture the variations in concepts.

\section{Conclusion} \label{sec:conclusion}
In this paper, we presented a data reuse aware co-optimization approach (DRACO) for improving the  PE utilization on systolic accelerators for memory-bound DNNs. We explained the role of computational complexity, PE utilization, and array size of the systolic accelerators on  (inference) latency optimization. We demonstrated that improving PE utilization does not always improve the performance of a DNN; it also depends on the computational overhead of improving PE utilization.  Our approach to improving PE utilization also boosts the predictive performance of DNNs.

{
\scriptsize
 \linespread{0.97}
\bibliographystyle{IEEEtran1}
\bibliography{References}

\begin{thebibliography}{10}
\providecommand{\url}[1]{#1}
\csname url@samestyle\endcsname
\providecommand{\newblock}{\relax}
\providecommand{\bibinfo}[2]{#2}
\providecommand{\BIBentrySTDinterwordspacing}{\spaceskip=0pt\relax}
\providecommand{\BIBentryALTinterwordstretchfactor}{4}
\providecommand{\BIBentryALTinterwordspacing}{\spaceskip=\fontdimen2\font plus
\BIBentryALTinterwordstretchfactor\fontdimen3\font minus
  \fontdimen4\font\relax}
\providecommand{\BIBforeignlanguage}[2]{{%
\expandafter\ifx\csname l@#1\endcsname\relax
\typeout{** WARNING: IEEEtran.bst: No hyphenation pattern has been}%
\typeout{** loaded for the language `#1'. Using the pattern for}%
\typeout{** the default language instead.}%
\else
\language=\csname l@#1\endcsname
\fi
#2}}
\providecommand{\BIBdecl}{\relax}
\BIBdecl

\bibitem{2017_Chen_JSSC}
Y.-H. Chen \emph{et~al.}, ``Eyeriss: An energy-efficient reconfigurable
  accelerator for deep convolutional neural networks,'' \emph{JSSC}, pp.
  127--138, 2016.

\bibitem{shao2019simba}
Y.~S. Shao \emph{et~al.}, ``Simba: Scaling deep-learning inference with
  multi-chip-module-based architecture,'' in \emph{MICRO}, 2019, pp. 14--27.

\bibitem{jouppi2017datacenter}
N.~P. Jouppi \emph{et~al.}, ``In-datacenter performance analysis of a tensor
  processing unit,'' in \emph{ISCA}, 2017, pp. 1--12.

\bibitem{chen2016eyeriss}
Y.-H. Chen \emph{et~al.}, ``Eyeriss: A spatial architecture for
  energy-efficient dataflow for convolutional neural networks,'' \emph{ACM
  SIGARCH Computer Architecture News}, vol.~44, no.~3, pp. 367--379, 2016.

\bibitem{Howard2017MobileNetsEC}
A.~G. Howard \emph{et~al.}, ``Mobilenets: Efficient convolutional neural
  networks for mobile vision applications,'' \emph{ArXiv}, vol. abs/1704.04861,
  2017.

\bibitem{2018_CVPR_Sandler}
M.~Sandler \emph{et~al.}, ``Mobilenetv2: Inverted residuals and linear
  bottlenecks,'' in \emph{CVPR}, June 2018, pp. 4510--4520.

\bibitem{Zhang_2018_CVPR}
X.~Zhang \emph{et~al.}, ``Shufflenet: An extremely efficient convolutional
  neural network for mobile devices,'' in \emph{CVPR}, 2018, pp. 6848--6856.

\bibitem{park2018deep}
J.~Park \emph{et~al.}, ``Deep learning inference in facebook data centers:
  Characterization, performance optimizations and hardware implications,''
  \emph{ArXiv'18}.

\bibitem{chen2019eyeriss}
Y.-H. Chen \emph{et~al.}, ``Eyeriss v2: A flexible accelerator for emerging
  deep neural networks on mobile devices,'' \emph{JETCAS}, pp. 292--308, 2019.

\bibitem{2019_DATE_Wu}
H.~{Wu} \emph{et~al.}, ``Data locality optimization of depthwise separable
  convolutions for cnn inference accelerators,'' in \emph{DATE}, 2019, pp.
  120--125.

\bibitem{2019_ASPDAC_Liu}
B.~Liu \emph{et~al.}, ``Addressing the issue of processing element
  under-utilization in general-purpose systolic deep learning accelerators,''
  in \emph{ASPDAC}, 2019, pp. 733–--738.

\bibitem{Fastai_Imagenette}
\url{https://github.com/fastai/imagenette}, 2019.

\bibitem{2017_DAC_Wei}
X.~Wei \emph{et~al.}, ``Automated systolic array architecture synthesis for
  high throughput cnn inference on fpgas,'' in \emph{DAC}, 2017, pp. 1--6.

\bibitem{2018_ICPR_Kung}
H.~T. {Kung} \emph{et~al.}, ``Adaptive tiling: Applying fixed-size systolic
  arrays to sparse convolutional neural networks,'' in \emph{ICPR}, 2018, pp.
  1006--1011.

\bibitem{gao2019tangram}
M.~Gao \emph{et~al.}, ``Tangram: Optimized coarse-grained dataflow for scalable
  nn accelerators,'' in \emph{ASPLOS}, 2019, pp. 807--820.

\bibitem{gao2017tetris}
M.~Gao \emph{et~al.}, ``Tetris: Scalable and efficient neural network
  acceleration with 3d memory,'' in \emph{ASPLOS}, 2017, pp. 751--764.

\bibitem{tan2019efficientnet}
M.~Tan \emph{et~al.}, ``Efficientnet: Rethinking model scaling for
  convolutional neural networks,'' in \emph{ICML}, 2019, pp. 6105--6114.

\bibitem{He_2016_CVPR}
He \emph{et~al.}, ``Deep residual learning for image recognition,''
  \emph{CVPR}, 2016.

\bibitem{guo2019depthwise}
Y.~Guo \emph{et~al.}, ``Depthwise convolution is all you need for learning
  multiple visual domains,'' in \emph{AAAI}, vol.~33, 2019, pp. 8368--8375.

\bibitem{Ioannou_2017_CVPR}
Y.~Ioannou \emph{et~al.}, ``Deep roots: Improving cnn efficiency with
  hierarchical filter groups,'' in \emph{CVPR}, 2017, pp. 1231--1240.

\end{thebibliography}
}

\end{document}